\documentclass[fleqn,usenatbib]{mnras}

\usepackage{newtxtext,newtxmath}

\usepackage[T1]{fontenc}

\DeclareRobustCommand{\VAN}[3]{#2}
\let\VANthebibliography\thebibliography
\def\thebibliography{\DeclareRobustCommand{\VAN}[3]{##3}\VANthebibliography}

\usepackage{graphicx}	
\usepackage{amsmath}	
\usepackage{nccmath}
\usepackage{multirow,booktabs,colortbl,tabularx}

\newcommand{\RNum}[1]{\uppercase\expandafter{\romannumeral #1\relax}}

\def\lcdm{$\Lambda$CDM}

\def\gassf{gas$_{\rm{SF}}$ }

\definecolor{Mygrey}{gray}{0.75}
\def\coco{co $\rightarrow$ co }
\def\cocounter{co $\rightarrow$ counter }
\def\counterco{counter $\rightarrow$ co }
\def\countercounter{counter $\rightarrow$ counter }
\def\trelax{$t_{\rm{relax}}$ }
\def\tdyn{$\bar{t}_{\rm{dyn}}$ }
\def\ttorque{$\bar{t}_{\rm{torque}}$ }
\def\ETGETG{ETG $\rightarrow$ ETG }
\def\LTGLTG{LTG $\rightarrow$ LTG }

\title[Timescales of kinematic misalignments]{Stellar-gas kinematic misalignments in \textsc{eagle}: lifetimes and longevity of misaligned galaxies}

\author[Maximilian K. Baker et al.]{
Maximilian K. Baker$^{1}$\thanks{E-mail: bakermk@cardiff.ac.uk},
Timothy A. Davis$^{1}$,
Freeke van de Voort$^{1}$,
Ilaria Ruffa$^{1,2,3}$
\\
$^{1}$Cardiff Hub for Astrophysics Research \&\ Technology, School of Physics \&\ Astronomy, Cardiff University, Queens Buildings, Cardiff, CF24 3AA, UK 
\\
$^{2}$INAF, Arcetri Astrophysical Observatory, Largo Enrico Fermi 5, I-50125 Florence, Italy
\\
$^{3}$INAF, Istituto di Radioastronomia, via P. Gobetti 101, 40129 Bologna, Italy
}

\date{Accepted 2025 June 9. Received 2025 May 30; in original form 2024 November 26}

\pubyear{\the\year{}}

\begin{document}
\label{firstpage}
\pagerange{\pageref{firstpage}--\pageref{lastpage}}
\maketitle

\begin{abstract}

The dominant processes by which galaxies replenish their cold gas reservoirs remain disputed, especially in massive galaxies. Stellar-gas kinematic misalignments offer an opportunity to study these replenishment processes. However, observed distributions of these misalignments conflict with current models of gas replenishment in early-type galaxies (ETGs), with longer relaxation timescales suggested as a possible solution. We use the \textsc{eagle} simulation to explore the relaxation of unstable misaligned gas in galaxies with masses of $M_{*}\geqslant \mathrm{10^{9.5}}$ M$_\odot$ between $0<z<1$. We extract misalignments from formation to relaxation providing a sample of $\sim3200$ relaxations. We find relaxation timescales tend to be short-duration, with median lifetimes of $\sim0.5$~Gyr, though with a notable population of unstable misalignments lasting $\gtrsim1$~Gyr. Relaxation time distributions show a log-linear relationship, with $\approx20$~per~cent of unstable misalignments persisting for $\gtrsim3$ torquing times. Long-lived unstable misalignments are predominantly found in galaxies with higher stellar masses, lower star-forming gas fractions, higher ongoing gas inflow, and which reside in the centres of dense environments. Mergers only cause $\approx10$~per~cent of unstable misalignments among galaxies at $z<0.35$, and $\approx21$~per~cent at $0.35<z<1.0$ in \textsc{eagle}. We conclude that, at least in \textsc{eagle}, unstable kinematic misalignments are not predominantly driven by gas-rich minor mergers at any redshift probed. Additionally, processes that significantly extend relaxation times are not dominant in the galaxy population. Instead, we see a diverse formation pathway for misalignments such as through hot halo cooling.

\end{abstract}




\section{Introduction}\label{intro:}

Throughout a galaxy's lifetime, the supply of gas plays an important role in driving its evolution. Cold, molecular gas in particular plays a key role as it is consumed by galactic processes such as star formation (SF) and active galactic nuclei (AGN) fuelling. However, with gas depletion timescales on the order of a few gigayears, systems must replenish their cold gas reservoirs if they are to maintain the observed levels of SF \citep[e.g.][]{Kennicutt94}.

Galaxies can replenish their cold gas reservoirs through a variety of processes. Internal replenishment can occur through stellar mass loss \citep[e.g.][]{ParriottBregman08, LeitnerKravtsov11} and the cooling of hot halo gas \citep[e.g.][]{Keres05, Lagos2014}. Cold gas can also be supplied externally through gas-rich galaxy interactions/mergers \citep[e.g.][]{DiTeodoro14, BarreraBallesteros2015} and direct accretion of cold gas from filaments \citep[e.g.][]{Keres05, Brooks09}. While the supply of cold gas through major mergers is fairly well established, the degree to which cold gas is replenished through other means is less constrained. 

Stellar-gas kinematic misalignments (henceforth `misaligned galaxies') have been one laboratory in which the origins of cold gas replenishment has been studied \citep[e.g.][]{Sarzi2006, Davis2011, DavisBureau2016, Bryant2019, Ruffa2019b, Duckworth2020a}. Observationally, misalignments are found in $\approx11$~per~cent of the galaxy population at $z\sim0.1$ \citep{Bryant2019} and show a strong morphological dependence, with $\sim30-40$~per~cent of ETGs and $\sim2-5$~per~cent of LTGs showing such features \citep[e.g.][]{Davis2011, Chen2016, Duckworth2019, Bryant2019, Ruffa2019a, Ristea2022, Raimundo2023}. Cosmological hydrodynamical simulations, such as \textsc{eagle} \citep[][]{SchayeCrain2015, Crain2015calibration} and \textsc{horizon-agn} \citep[][]{Dubois14}, have also proven to be invaluable to understanding the physical drivers and timescales involved in misalignments, owing to their ability to trace galaxies over time. These have been able to recreate the observed misalignment populations \citep[e.g.][]{Duckworth2020a, Khim2020, Casanueva2022}. 

The existence of stellar-gas misalignments is predicted by our current Lambda Cold Dark Matter ($\Lambda$CDM) cosmological model. Stellar and gaseous components are expected to form an aligned discy structure as a natural consequence of galaxy formation \citep[see tidal torque theory; TTT by][]{Peebles69, Doroshkevich70, White84}. As the galaxy evolves, stellar and gas components may decouple and become misaligned, provided that the angular momentum of the existing gas disc is overcome and a new, misaligned disc can form. The stochastic nature of external gas accretion and ability, to first order, to accrete with a random angular momentum orientation has led external accretion to become the favoured dominant driver behind many misalignments \citep[e.g.][]{Davis2011, VoortDavis2015, Starkenburg2019, Duckworth2020b, Khim2021, Casanueva2022}.

Due to their inherent instability, stellar-gas kinematic misalignments are transient phenomena and are not expected to be long-lived \citep[][]{Tohline1982, LakeNorman1983}. Once formed, a misaligned gas disc will continuously dissipate angular momentum until it has settled back into the plane of the stellar disc (assumed to dominate the mass distribution at radii of interest). This produces populations of conventional co-rotating $(\sim0^{\circ})$ and counter-rotating $(\sim180^{\circ})$ galaxies. While the former are indistinguishable from secularly evolved galaxies, the latter are long-lived and thought only possible through the relaxation of a previously misaligned gas disc \citep[e.g.][]{Bryant2019, Duckworth2020a, Khoperskov2020}. Prolate mass distributions with a polar-ring $(\sim90^{\circ})$ are also thought to be stable \citep[][]{Tohline1982}, though such systems are not common and constitute only a few percent of observable lenticular galaxies \citep[e.g.][]{MaccioMooreStadel06, Bryant2019}. In the absence of processes preserving this misaligned gas disc, relaxation into a stable regime is thought to occur quickly, with idealised theoretical estimates predicting timescales of less than 5 dynamical times \citep[][]{Tohline1982, SteimanCameron88}. For a typical ETG with a highly spherical mass distribution, this is typically equivalent to $\sim100$ Myr \citep[][]{DavisBureau2016}. 

A key area of contention lies between the observed shape of the misalignment fraction distributions and those predicted by analytical models. As explored in detail in \citet[][]{DavisBureau2016}, assuming mergers are the primary driver of misalignments alongside theoretical relaxation timescales (in the absence of smooth accretion), current models significantly overpredict the number of relaxed systems. Recent theoretical findings point to a significantly more diverse formation path for misaligned systems. These include the accretion of gas stripped from a nearby low-mass satellite, perturbations of the existing co-rotating gas induced from fly-bys or ram-pressure stripping, and halo cooling onto gas-poor systems \citep[e.g.][]{Lagos2015, Bryant2019, Khim2021, Cenci2023}. Likewise, the inclusion of smooth accretion onto an existing misaligned gas disc has been shown to significantly enhance the misalignment timescales beyond theoretical timescales \citep[e.g.][]{VoortDavis2015, Lagos2015, Ruffa2019b}. This misaligned smooth accretion may be supplied through halo cooling \citep[e.g.][]{Negri2014, Lagos2015, Ruffa2019b, Duckworth2020a}, provided the galaxy hosts a gaseous halo that is kinematically misaligned with respect to its stellar component \citep[e.g.][]{Padilla2014, Duckworth2020a}. Misaligned smooth accretion may also be supplied from tidal tails from a recent merger which can significantly extend the longevity of a misalignment beyond $>2$~Gyr \citep[][]{VoortDavis2015}. Taken together, these results suggest the processes by which galaxies replenish their cold gas may be more diverse than previously thought. Thus, by investigating the timescales of misalignments in hydrodynamical simulations we can gain insights into gas replenishment processes that may be challenging to detect observationally. 

Thus far, the exploration of misalignment timescales has been largely limited to small numbers of galaxies with idealised behaviour. One notable exception is from \citet{Khim2021} who used the \textsc{horizon-agn} simulation to study the lifetimes of misalignments in a sizeable galaxy population. Using a sample of $\sim2200$ misaligned galaxies at $z=0.52$ and measuring the population decay, they find a significant population of galaxies with misalignment lifetimes of $>2$~Gyr with a strong morphological dependence. ETGs were found with characteristic decay timescales of $\approx2.3$~Gyr compared to $\approx0.8$~Gyr for LTGs. Overall, galaxies with lower gas fractions, more early-type morphologies, and those residing in denser environments were found to have longer decay timescales. As these properties are common among ETGs \citep{Dressler80}, this is thought to amplify the higher misalignment fractions among ETG populations. Additionally, misalignments formed from interactions between the galaxy and a dense medium \citep[e.g. ram-pressure stripping; ][]{GunnGott72} were found to be the most long-lived, with those caused by mergers, galaxy interactions, and remaining processes including halo cooling and filament accretion resulting in faster relaxations. 

Another noticeable exception is from \citet{Casanueva2022} who used the \textsc{eagle} simulation to investigate the origin of misaligned galaxies for a representative galaxy population. They find galaxies that remain misaligned between two neighbouring simulation snapshots are typically more spheroidal, dispersion-dominated, and gas-poor. Misalignment formation was found to correlate most strongly with a reduction in cold gas fraction, which was attributed to gas stripping from nearby galaxy interactions. Mergers were found to coincide with misalignment formation in only $\approx4$~per~cent. Alongside comparable external accretion rates in aligned galaxies, \citet{Casanueva2022} suggest external processes are not the dominant driver of misalignments in \textsc{eagle}, in agreement with \citet{Khim2021}.

In this paper, we use the \textsc{eagle} (Evolution and Assembly of GaLaxies and their Environments) suite of cosmological hydrodynamical simulations \citep{SchayeCrain2015, Crain2015calibration} to investigate the persistence of stellar-gas kinematic misalignments for a representative galaxy population from $0<z<1$. Importantly, our approach differs from that of \citet{Khim2021} as we extract a population of misalignments over a complete window from formation to relaxation, rather than characterising relaxation timescales using a misalignment population decay from a given redshift. Additionally, we study these misalignments as 3D angles, rather than in projection, and treat stable co- and counter-rotating regimes separately. In this way, we create a reliable estimate of the misalignment timescales for a large galaxy sample for the first time. Specifically, we aim to answer the following questions: 1) whether significantly longer relaxation timescales (of order $\sim20-80$ dynamical timescales) as suggested by \citet{DavisBureau2016} are seen within \textsc{eagle}; 2) the degree to which relaxation times vary with galaxy properties and environment; and 3) to what extent are mergers associated with misalignment formation. 

This paper is organised as follows. In Section~\ref{theory:} we outline the theoretical background regarding predicted relaxation timescales. In Section~\ref{EAGLE:} we give an overview of \textsc{eagle}. In Section~\ref{sample:} we outline the sample of galaxies we use for this work. In Section~\ref{method:} we give an overview of the galaxy properties we calculate alongside definitions of various parameters we use throughout this work. In Section~\ref{results:} we present our results and compare these with existing results. Finally, we conclude in Section~\ref{conclusions:}.


\section{Theoretical relaxation timescales}\label{theory:}

As outlined by \citet{Tohline1982}, a misaligned disc of gas can be modelled by a series of rings embedded within an ellipsoidal gravitational potential well. These rings of gas have a circular velocity $V_{\mathrm{rot}}$ at a radius $R$. The non-zero angular momentum component of this gas will experience a net torque from the existing mass distribution. This results in a precession rate given by 
\begin{ceqn}
    \begin{align}\label{eq: precession rate}
    \dot{g} = -\frac{3}{2}\omega_{0}J_{2}\mathrm{cos}(\psi),
    \end{align}
\end{ceqn}
where $\omega_{0} = V_{\mathrm{rot}}/R$ is the angular velocity of the gas at a given radius, $J_2$ is the coefficient of the quadrupole moment of an ellipsoidal gravitational field, and $\psi$ is the angle between the stellar and cold gas angular momentum vectors.

As the precession rate given by equation~\ref{eq: precession rate} increases toward smaller radii, this results in a warped disc. This planar divergence is stronger toward the central regions of the disc and has been well-studied in nearby galaxies such as Centaurus A \citep[e.g.][]{Nicholson92}. For a misaligned gas disc, neighbouring non-coplanar cold gas clouds will interact and dissipate the perpendicular angular momentum component. This suppresses non-circular orbits and forces radially adjacent rings into co-planar orbits. The angular momentum of a misaligned gas disc is therefore not conserved, with the surrounding stars absorbing the lost angular momentum \citep{Tohline1982}. The loss of angular momentum causes the gas disc to settle into the fundamental plane of the galaxy given by the angular momentum of the mass distribution. As the mass of stars is thought to dominate the mass distributions of the centres of these galaxies, the fundamental plane is assumed to be that of the net stellar angular momentum. A misaligned gas disc is therefore expected to naturally settle into a co- or counter-rotating configuration with respect to the stellar component. As there are no net torque acts on gas discs in co- or counter-rotating configurations, these are expected to be long-lived and stable.

As described by \citet{Tohline1982} and \citet{LakeNorman1983}, in the absence of continued accretion onto the gas disc that may preserve its angular momentum, the time taken for a misaligned disc to relax back into a stable configuration should be governed by the torquing timescale, 

\begin{ceqn}
    \begin{align}\label{eq: relax time simple}
    t_{\mathrm{torque}} \approx \frac{1}{\dot{g}} \approx \frac{t_{\mathrm{dyn}}}{\epsilon} = \frac{2\pi R}{V_{\mathrm{rot}}\epsilon},
    \end{align}
\end{ceqn}
where $\epsilon$ is the stellar ellipticity and the dynamical time of the galaxy is defined as $t_{\mathrm{dyn}}=2\pi R/V_{\mathrm{rot}}$. For a typical ETG with $\epsilon \approx 0.2$, a disc size of $R\approx1$ kpc, and $V_{\mathrm{rot}}\approx200$ km s$^{-1}$, this implies $t_{\mathrm{relax}} \approx 5t_{\mathrm{dyn}}$, corresponding to $\sim 0.2$~Gyr. This highlights the expected short-lived lifetimes of stellar-gas kinematic misalignments.


\section{The \textsc{eagle} simulation}\label{EAGLE:}

The Evolution and Assembly of GaLaxies and their Environments\footnote{Publicly available data products available from \url{http://eagle.strw.leidenuniv.nl} and \url{https://icc.dur.ac.uk/Eagle} and described in \citet{Mcalpine2016}} (\textsc{eagle}) is a suite of cosmological hydrodynamical simulations described in \citet{SchayeCrain2015} and \citet{Crain2015calibration}. These simulations use a modified version of the parallel N-body smoothed particle hydrodynamics (SPH) code \textsc{Gadget}-3 \citep{Springel05, Springel2008} known as \textsc{Anarchy} (see \citealt{Schaller2015} for description and analysis). The initial conditions for these simulations were generated using \textsc{Panphasia} \citep{JenkinsBooth13}. A \lcdm~framework is also assumed, with $\Omega _{\rm{m}} = 0.307$, $\Omega _{\rm{\Lambda}} = 0.693$, $\Omega _{\rm{b}} = 0.04825$, $h = 0.6777$, $\sigma _{\rm{8}} = 0.8288$, $n_{\rm{s}} = 0.9611$ and $Y = 0.248$ \citep{PlanckCollaboration14}.

\textsc{eagle} uses sub-grid models to capture physical processes below the resolution limit ($\sim0.7$ pkpc) of the simulation. These include radiative cooling and photoheating \citep{Wiersma09a}, star formation \citep{SchayeDallaVecchia08}, stellar evolution and interstellar medium (ISM) enrichment \citep{Wiersma09b}, the stochastic thermal stellar feedback by \citet{DallaVecchiaSchaye12}, and black hole growth through accretion and mergers and feedback \citep{Springel05, RosasGuevara2015}. The \textsc{eagle} sub-grid parameters are tuned to reproduce observations of the galaxy stellar mass function and galaxy sizes relation at $z\sim0$. The \textsc{eagle} simulations have been successful in reproducing observational mass, size, and morphology distributions \citep[e.g.,][]{SchayeCrain2015, Correa2017, TrayfordSchaye2018, Pfeffer2023}.

\begin{table}
	\centering
	\caption{Parameters of the \textsc{Ref-L100N1504} simulation from the \textsc{eagle} suite. From top to bottom, the rows report: co-moving box size; initial dark matter (DM) particle count (initially there is an equal number of baryonic particles); initial particle mass of gas particles; initial particle mass of DM particles; co-moving, Plummer-equivalent gravitational softening length; maximum physical gravitational softening length.}
	\label{table:simulation_parameters}
	\begin{tabular}{lcr} 
		\hline
        Parameter & Value\\
		\hline
        Volume & $(100$ cMpc$)^3$\\
        total particle count & $\times 1504^3$\\
        Initial gas particle mass & $1.81 \times 10^6$ M$_\odot$\\
        Initial DM particle mass & $9.7 \times 10^6$ M$_\odot$\\
        Co-moving gravitational softening length & 2.66~ckpc\\
        Max. physical gravitational softening length & 0.7~pkpc\\
		\hline
	\end{tabular}
\end{table}

In \textsc{eagle}, galaxies and their host halos are identified through a series of stages described in \citet{SchayeCrain2015}. In short, dark matter halos are identified in post-processing using the friend-of-friends (FoF) algorithm \citep{Davis85}, with a linking length of $\approx 0.2$ times the mean interparticle separation. Stellar, gas, and black hole particles are then assigned to the same FoF-halo as their nearest DM particle. Substructures within these FoF-halos, called subhalos, are then identified using the \textsc{subfind} algorithm \citep{Springel01, Dolag09} by identifying gravitationally-bound overdensities by considering both DM and baryonic particles. The subhalo with the lowest value of the gravitational potential within a FoF-halo is assigned as the central galaxy, with the remaining subhalos being designated as satellite galaxies. 

Galaxy evolution can be traced using the subhalo merger trees by \citet{Qu2017} that were generated using the \textsc{D-Trees} algorithm \citep{Jiang14}. The latter links subhalos by establishing the location of the majority of all particles assigned to a particular subhalo between consecutive snapshots\footnote{For a more detailed explanation of the \textsc{D-Trees} algorithm and merger tree construction see \citet{Jiang14} and \citet{Qu2017}, respectively}. Mergers are treated hierarchically such that two (or more) progenitor galaxies undergoing a merger will form a single descendant galaxy. A merger is defined to occur once the \textsc{D-Trees} algorithm can no longer identify a progenitor subhalo among the \textsc{subfind} catalogues. At this point the \textsc{subfind} algorithm can no longer distinguish the progenitor density profiles, at which point particles from the secondary galaxy are reassigned to the primary galaxy. Occasionally during mergers, the algorithm may switch the primary and secondary galaxies if these are in very close proximity \citep[`Subhalo switching problem'; see][for details]{Qu2017}. Regarding this work, a switch between coalescing galaxies in which one galaxy is misaligned may distort the measured misalignment duration. To minimise this effect, we exclude galaxies that experience a spontaneous stellar mass decrease of $>50$~per~cent whilst misaligned. This ensures a spontaneous switch to a low-mass merging satellite is not falsely included as a misalignment. However, we do not find this choice to meaningfully affect our sample or results.

The \textsc{eagle} simulation is stored in 400 intervals between $z\sim20$ to $z=0$. Of these, 29 are "snapshots" with time intervals from $\approx 0.3$ to 1~Gyr, which record complete particle properties alongside extensive halo and subhalo property catalogues. Additionally, 400 higher time-cadence "snipshots" are stored with limited catalogues of particle and subhalo properties, though with enough information to calculate misalignment angles. We use 200 of these 400 snipshots for which merger trees were run by \citet[][]{Crain2017}, with a mean time cadence of $\approx0.12$~Gyr within our target redshift range. These allow us to distinguish between faster ($\sim0.1$~Gyr) and longer-lived ($\gtrsim1$~Gyr) misalignments. As we wish to focus on the period in which galaxy growth occurs largely linearly, we limit our data analysis to $0<z<1$ which provides a wide temporal window over which to measure relaxation timescale distributions. For this work we make use of the largest reference simulation, \textsc{Ref-L100N1504}, which has a box size of 100 co-moving Mpc$^3$ (henceforth ckpc for co-moving and pkpc for proper length-scales). Parameters specific to the \textsc{Ref-L100N1504} are summarised in Table~\ref{table:simulation_parameters}. 


\begin{figure}
    \begin{center}
    \includegraphics[width=8.5cm]{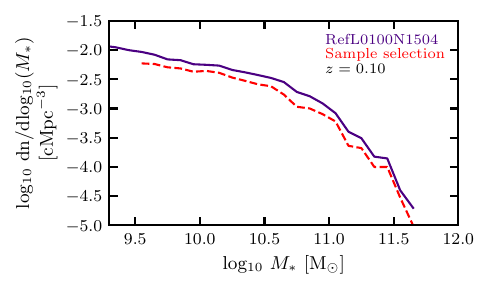}\\
    \caption{Galaxy stellar mass function at $z=0.1$ of the total galaxy population (solid purple line) in the \textsc{Ref-L100N1504} simulation in comparison with our sub-sample of galaxies (dashed red line) selected using the criteria listed in Section~\ref{sample:}. There is no mass bias in our sub-sample compared to the general galaxy population.}
    \label{plot: stellarMassFunc}
    \end{center}
\end{figure}

\section{Sample}\label{sample:}

The \textsc{Ref-L100N1504} simulation contains $\sim7,300$ galaxies with a stellar mass $M_{*}>10^{9.5}$~M$_{\odot}$ at $z=0.1$. This provides a statistically-significant galaxy population with a range of morphologies and sizes from which to sample misalignments. 

In order to select our sample we use similar limits as those adopted by \citet{Casanueva2022} and highlight any differences below. We consider both central and satellite galaxies in our analysis and extract galaxies under the following conditions:

\begin{enumerate}
    \item A minimum subhalo stellar mass of $M_{*}>10^{9.5}$~M$_{\odot}$ within a spherical aperture size of 30~pkpc, centred on the minimum potential of the galaxy. This mass limit ensures we only include galaxies that are sufficiently resolved.
    \item A minimum star-forming gas (gas$_{\rm{SF}}$) particle count of 20 for kinematic measurements within the stellar half-mass radius ($r_{50}$). We define $r_{50}$ as the spherical radius enclosing half the stellar mass within 30 pkpc, centred on the stellar centre of mass, which is approximately equivalent to the effective radius in galaxy observations. The low gas$_{\rm{SF}}$ particle count allows us to probe relatively deep into the low quenched ETG population (that dominates the misaligned population) whilst having a statistically significant particle count from which to extract kinematic information. We note that changing this limit up to 50 or 100 particles leaves the shape and median values of our relaxation timescale distributions unchanged, and thus this choice does not unduly influence our results.
    \item A maximum spatial separation between the stellar and gas$_{\rm{SF}}$ centres of mass within $<2$~pkpc. This allows us to exclude galaxies featuring highly perturbed cold gas reservoirs that are only loosely associated with the galaxy \citep[see Appendix D in][for examples of this in high-mass galaxies]{TrayfordSchaye2018}.
    \item A maximum uncertainty of $<30^{\circ}$ for the misalignment angle using boostrap resampling.  This criterion is motivated by the presence of systems that are heavily dispersion-dominated and/or feature relatively few gas$_{\rm{SF}}$ particles (as is typical for many ETGs). Changing this limit to $<10^{\circ}$ leaves our results and conclusions unchanged.
\end{enumerate}

Unlike \citet{Casanueva2022}, we do not impose a direct limit on dispersion-dominated systems through the use of the stellar spin parameters \citep[see][]{Emsellem2011FRSR}. Instead, we use bootstrap resampling and take the $1\sigma$ ($68$~per~cent) percentiles to estimate the uncertainty in the misalignment angle.  This uncertainty method will naturally scale with the degree of stellar dispersion in the galaxy, but will also remove a handful of galaxies with chaotic \gassf kinematics that are not captured by the stellar spin parameter. Our uncertainty limit of $<30^{\circ}$ is motivated by similar uncertainty thresholds used in observational studies \citep[e.g.][]{Davis2011}. At $z=0.1$, this criterion reduces our sample size by $\approx6.5$~per~cent. The majority of the excluded objects are dispersion-dominated ETGs with low \gassf particle counts. 

At $z=0.1$, this gives us a sample of 5428 galaxies. Similar numbers are obtained at redshifts up to $z\approx1$. Overall, the adopted criteria produce galaxy samples that do not show significant bias towards any galaxy mass, for a range of snapshots. For instance, Figure~\ref{plot: stellarMassFunc} shows the stellar mass function of our final sample at $z=0.1$, alongside the total galaxy population in the \textsc{Ref-L100N1504} reference simulation, which shows that the two populations have a similar mass distribution down to our sample limit. Additionally, this sample shows no bias in $M_*-$SFR (among galaxies with SFR~$>0$) and no bias in halo mass (not shown).

\section{Methodology}\label{method:}

\subsection{Internal galaxy properties}\label{method: internal properties}

\subsubsection{Star-forming gas and star-forming gas fraction}

Observationally, gas kinematics are typically measured by tracing the warm interstellar medium using H$\alpha$. This has been shown to be co-moving with the cold dense phase of the ISM \citep[e.g.][]{Davis2011}. For this work, we assume that the cold gas in a galaxy is described most closely by the properties of the star-forming gas (gas$_{\rm{SF}}$) particles in \textsc{eagle}. These are defined as gas particles with a SFR $>0$ and follow a metallicity-dependant density threshold described in \citet{SchayeCrain2015}. 

We define the star-forming gas fraction of a galaxy as 
\begin{ceqn}
    \begin{align}
 f_{\rm{gas,SF}} = \frac{M_{\rm{gas,SF}}}{M_{\rm{gas}}+M_{*}},
    \end{align}
\end{ceqn}
where $M_{*}$, $M_{\rm{gas}}$, and $M_{\rm{gas,SF}}$ are the stellar, gas, and gas$_{\rm{SF}}$ masses of the galaxy within our kinematic aperture of $r_{50}$. 

\subsubsection{Morphology}

\textsc{eagle} successfully reproduces the observed bimodal galaxy population of `red sequence' spheroidal quiescent galaxies (ETGs) and `blue cloud' discy star-forming galaxies \citep[LTGs; e.g.][]{Trayford2015, Correa2017, Thob2019, Pfeffer2023}.

\citet{Correa2017} found that these populations can be effectively distinguished using the ratio of the co-rotational kinetic energy fraction, $\kappa_{\rm{co}}$, as a proxy for morphology. This describes the ratio of total kinetic energy ($K$) that is invested in co-rotation ($K^{\rm{rot}}_{\rm{co}}$) and is given by
\begin{ceqn}
    \begin{align}
    \kappa _{\rm{co}} = \frac{K_{\rm{co}}^{\rm{rot}}}{K} = \frac{1}{K}\sum_{i, L_{\rm{z}, \textit{i}} > 0} \frac{1}{2}m_{i}\left ( \frac{L_{\rm{z}, \textit{i}}}{m_i R_i} \right )^{2},
    \end{align}
\end{ceqn}
where the sum is over all particles of a given type within a sphere of radius 30 pkpc centred on the minimum potential, $m_{i}$ is the particle mass, $L_{\rm{z},\textit{i}}$ is the net angular momentum, and $R_{i}$ is the radius from the centre of potential in the plane normal to the rotation axis of the given particle type \citep{Correa2017}. We denote stellar co-rotational kinetic energy fraction as $\kappa_{\rm{co}}^*$. Following \citet{Correa2017} and \citet{Thob2019}, we use $\kappa_{\rm{co}}^*$ as a proxy for morphology to distinguish between ETGs ($\kappa_{\rm{co}}^* < 0.4$) and LTGs ($\kappa_{\rm{co}}^* > 0.4$).

Similarly, we define the \gassf co-rotational kinetic energy fraction as $\kappa_{\rm{co}}^{\rm{SF}}$ using all \gassf particles within a sphere of radius $r_{50}$. We find by visual inspection that thin gas discs can be broadly described by $\kappa_{\rm{co}}^{\rm{SF}}\gtrsim0.7$, which is consistent to existing literature \citep[e.g.][]{Jimenez2023}.

To characterise the torquing timescale, we make use of a further proxy of galaxy morphology, the stellar ellipticity, using the iterative scheme outlined in \citet{Thob2019}. We define the stellar ellipticity as
\begin{ceqn}
    \begin{align}
    \epsilon = 1 - \frac{c}{a},
    \end{align}
\end{ceqn}
where $a$ and $c$ are the major and minor axes of the ellipse, respectively. As opposed to $\kappa_{\rm{co}}$, these use the square root of the eigenvalues of the stellar mass distribution tensor to characterise the morphology of a galaxy within an initial spherical aperture of 30 pkpc \citep[for details see][]{Thob2019}.

\subsubsection{Misalignment angles}

We assume the kinematic centre of the galaxy is described by the stellar centre of mass and stellar peculiar velocity within $r_{\rm{50}}$. We additionally impose $r_{\rm{50}} = \rm{max}(r_{\rm{50}}, 1\:\rm{pkpc})$ for both spherical and projected radii to account for artefacts at or near the resolution limit of $\approx0.7$~pkpc. 

We assume the kinematics of the particles are captured by the net angular momentum vector. For galaxies meeting our selection criteria, the angular momentum of a component is given by 
\begin{ceqn}
    \begin{align}
    \mathbf{L_{\rm{spin}}} = {\sum _{i} m_{i} \left ( \mathbf{r}_{i} - \mathbf{r}_{\rm{COM}} \right ) \times \left ( \mathbf{v}_{i} - \mathbf{v}_{\rm{COM}} \right )},
    \end{align}
\end{ceqn}
where the sum is over all particles of a specified type within $r_{50}$, $m_{i}$ is the particle mass, $\mathbf{r}_{i}$ and $\mathbf{v}_{i}$ are the physical position and velocity of the particle, and $\mathbf{r}_{\rm{COM}}$ and $\mathbf{v}_{\rm{COM}}$ are the position and peculiar velocity of the stellar centre of mass within $r_{50}$ which were used to ensure consistency with observational studies. For dark matter (DM) particles, we extract kinematic properties within a spherical aperture of 30 pkpc.

As we intend to investigate the inherent behaviour and evolution of misaligned disc kinematics, we predominantly work with misalignment angles in three-dimensional (3D) space, as opposed to projected misalignments between position angles (PAs) that are typically used in observational studies. This allows us to measure the relaxation timescales removing the bias of the viewing angle. Misalignment angles are given by the dot product between the angular momentum unit vectors. 

We assume the most widely used classification to interpret galaxy kinematics \citep[e.g.][]{Davis2011}, classifying as aligned those galaxies with stellar-gas kinematic angles $<30^{\circ}$, misaligned those with angles $>30^{\circ}$. Among the misaligned galaxy population, we make a further distinction between \textit{unstable} misalignments with angles $[30^{\circ}-150^{\circ}]$ and counter-rotation with angles $>150^{\circ}$.

\subsection{External properties}\label{method: external}

Following \citet{Casanueva2022}, we use the halo mass, $M_{\rm{200c}}$, as a proxy to describe the large-scale environment density within which a galaxy resides (e.g. galaxy clusters with $M_{\rm{200c}}>10^{14}$~M$_{\odot}$). $M_{\rm{200c}}$, is given by the total mass bound within a radius 200 times the critical density. Among central galaxies, it follows that a larger halo mass corresponds to a larger reservoir of hot halo gas available to potentially cool onto the central galaxy \citep[e.g.][]{Correa2018haloacc, MitchellSchaye2022halo}.

We define galaxy properties related to mergers using the hierarchical merger trees by \citet{Qu2017}. Galaxies with more than one progenitor galaxy is indicative of a galaxy merger having occurred. We define the mass ratio of a merger as 
\begin{ceqn}
    \begin{align}\label{eq: merger fraction}
    \mu_{\rm{*}} = \frac{M^{S}_*}{M^{P}_*},
    \end{align}
\end{ceqn}
where ${M^{S}_*}$ and $M^{P}_*$ are the stellar masses of the secondary (less massive) and primary (most massive) galaxies, respectively. Merger ratios are calculated from values in the merger tree at the point of the maximum stellar mass of the secondary galaxy within the past 2~Gyr to account for stellar stripping prior to infall \citep[see][]{RodriguezGomez15, Qu2017}. Major and minor mergers are defined with mass ratios of $\mu_{\rm{*}} \geq 0.3$ and $0.3 \geq \mu_{\rm{*}} \geq 0.1$, respectively \citep[e.g.][]{Cox08, Lagos2018a}. We assume that ratios of $\mu_{\rm{*}} < 0.1$ are direct accretion events (sometimes called micro-mergers) because these are poorly resolved in \textsc{eagle} \citep[][]{Crain2017}. 

\subsection{Relaxation times and relaxation paths}\label{method: Misalignments and relaxations}

\subsubsection{Relaxation times}\label{method: relaxation time definition}

In order to estimate the relaxation time of an unstable misalignment we require galaxies to meet our sample criteria over consecutive snapshots from misalignment formation to relaxation. We also exclude unstable misalignments for which we cannot establish a formation and relaxation point within $0<z<1$. By limiting ourselves to galaxies with a complete `relaxation window', we extract a smaller sample of relaxations but are able to establish a more reliable estimate of relaxation times. Additionally, our redshift range naturally places an upper limit on the relaxation time that we can accurately measure in our sample, with $t_{\mathrm{relax}}\approx40t_{\mathrm{dyn}}$ approaching the $\approx8$~Gyr window between $0<z<1$, for a typical ETG. 

In previous work using a single galaxy in a zoom-in simulation, \citet{VoortDavis2015} define the relaxation time as the time between the formation of the misalignment (an initial sharp peak to $\psi_{\rm{3D}}\approx90^{\circ}$) and the first snapshot with $\psi_{\rm{3D}}<20^{\circ}$. In contrast, \citet{Khim2021} use a sample of 2177 misalignments with projected PAs at $z=0.52$ to define the misalignment duration by fitting the remaining fraction of misaligned discs as a function of time. Galaxies are assumed to have relaxed upon first returning back to $\psi_{\rm{2D}}\leq30^{\circ}$. While this latter approach is able to mimic observational studies more reliably through the use of PAs, it is explicitly a measure of the duration a galaxy spends misaligned ($\psi_{\rm{2D}}\geq30^{\circ}$), rather than the duration a galaxy spends in the unstable misaligned regime. As such, this method does not account for systems that remain in the kinematically stable counter-rotating regime for a considerable time \citep[e.g.][]{Starkenburg2019}.

For this work, we adopt a similar definition and approach as \citet{VoortDavis2015}, modified for use on a more diverse population of galaxies and misalignments. Our aim here is to generalise relaxation timescales of unstable misalignments for comparison with theoretical predictions (see Section~\ref{theory:}), complementing the approach of \citet{Khim2021}. As an unstable misalignment may be formed from an existing long-lived counter-rotating system, we define a separate kinematically-stable regime of $\psi_{\rm{3D}}<20^{\circ}$ and $\psi_{\rm{3D}}>160^{\circ}$. This is motivated by visual inspection of typical galaxies that are found to mostly reside within this regime, with only few outliers caused by random fluctuations. Galaxies with $20^{\circ} \leq \psi_{\rm{3D}} \leq 160^{\circ}$ are defined as kinematically-unstable. We then require any galaxy leaving the kinematically-stable regime to enter the unstable \textit{misaligned} regime ($30^{\circ}\leq \psi_{\rm{3D}} \leq 150^{\circ}$) in order to be classified as significant deviation from the kinematically-stable regime\footnote{Following our existing method, we note that changing the stable regime limits to $\psi_{\rm{3D}}<30^{\circ}$ and $\psi_{\rm{3D}}>150^{\circ}$, and the unstable misaligned regime to $45^{\circ} \leq \psi_{\rm{3D}} \leq 135^{\circ}$ does not meaningfully change our results.}. In most cases, the first snipshot in the unstable regime also corresponds to the first snipshot in the unstable misaligned regime. We define the beginning of a relaxation from the first snipshot in the kinematically unstable regime.

As some misalignments may experience a small degree of scatter during relaxation to the stable regime, we introduce a new parameter, the latency time, $t_{\rm{lat}}$. This is defined as the minimum amount of time a galaxy must have spent in the stable regime for the unstable misalignment to have truly settled. We set this value to $t_{\rm{lat}}=0.1$~Gyr ($\gtrsim1$ snipshots), again from visual inspection. We define the end of a relaxation as the first snipshot upon returning to the stable regime, so long as the galaxy remains in the stable regime for period of at least another $t_{\rm{lat}}$. The relaxation time, $t_{\rm{relax}}$, is then defined as the time between the snipshots corresponding to the beginning and end of the misalignment. This gives us an error of approximately $\pm0.12$~Gyr on $t_{\rm{relax}}$.

By applying this criterion, we aim to find a balance between extracting numerous short-duration unstable misalignments that may be associated with small fluctuations from a larger overall relaxation, and extracting clearly separate unstable misalignment events as one long relaxation. We demonstrate the reliability of this criterion by showing two example galaxies in Section~\ref{results: galaxy evolutions} and discuss its implications on our relaxation timescale distribution in Appendix~\ref{Appendix: latency time}.

\subsubsection{Dynamical and torquing times}\label{method: tdyn ttorque definition}

Estimates of the $t_{\rm{dyn}}$ and $t_{\rm{torque}}$ require an estimate of the radius of the \gassf disc. We assume that this radius is broadly described by twice the \gassf half-mass radii ($2r^{\mathrm{SF}}_{\mathrm{50}}$) per visual inspection of our sample and from past studies on SF gas discs in \textsc{eagle} \citep[e.g.][]{Hill2021}. We calculate the rotational velocity ($V_{\mathrm{rot}}$) assuming a spherical mass distribution at a radius $\mathrm{min}(2r^{\mathrm{SF}}_{\mathrm{50}}, r_{\mathrm{50}})$ where we have taken into account the stellar half-mass radius ($r_{50}$) as the aperture of kinematic measurements. This allows us to compensate for galaxies with small gas discs with respect to the aperture of kinematic measurements, as may be found in some ETGs.

The $t_{\rm{torque}}$ of a galaxy can vary between snipshots due to changes in the gas disc radius and stellar ellipticity. The gas disc radius and stellar ellipticity were found to vary within a standard deviation of $\sim10$~per~cent between misalignment formation and relaxation in our sample. We express the relaxation time in terms of time passed in~Gyr and as normalised values using the the time-averaged dynamical and torquing time taken over the duration of the misalignment, denoted as \tdyn and \ttorque, respectively. Following theoretical predictions (see Section~\ref{theory:}), normalised relaxation timescales above $t_{\rm{relax}}/\bar{t}_{\rm{torque}}\gtrsim1$ would be indicative of processes acting to preserve the duration of the misalignment. To account for approximations made, we define long-duration unstable misalignments as those with $t_{\rm{relax}}/\bar{t}_{\rm{torque}}>3$.

\subsubsection{Relaxation paths}\label{method: relaxation paths}

Because we define relaxation timescales with respect to galaxies leaving and re-entering the co- and/or counter-rotating regimes, it follows that some unstable misalignments may transition between differing stable regimes. We exclude the meta-stable polar-ring orientation as one of the stable regimes. Indeed, although longer-lived than most misalignment configurations, these are thought to eventually relax into one of the aforementioned long-lived regimes \citep{Tohline1982, LakeNorman1983}. However, because galaxies that transition through this meta-stable regime may be expected to experience longer relaxation times, we choose to distinguish between relaxation paths.

For the majority of our sample, galaxies begin and end in a co-rotating configuration, denoted as co $\rightarrow$ co. However, we also see a variety of other relaxation pathways leading to populations of co-rotating to counter-rotating (co $\rightarrow$ counter), and more rarely counter-rotating to co-rotating (counter $\rightarrow$ co) and counter-rotating to counter-rotating (counter $\rightarrow$ counter) relaxation paths.

\subsubsection{Internal galaxy properties during relaxation}\label{method: properties during misalignment}

Unless stated otherwise, we approximate the internal properties of a galaxy by taking time-averaged values over the duration of the relaxation time. For morphological classifications that may be poorly captured by the time-averaged ${\kappa}_{\rm{co}}^*$, we instead consider its morphology pre- and post-misalignment. We take the average ${\kappa}_{\rm{co}}^*$ for the first $0.1$~Gyr pre- and post-misalignment. We can therefore classify galaxies as a mixture of those that retained their morphological classification before and after misalignment (denoted as e.g. ETG $\rightarrow$ ETG), and those that morphologically transitioned (denoted as e.g. LTG $\rightarrow$ ETG).


\section{Results}\label{results:} 

\subsection{Misalignment distributions}\label{results: misalignment distributions}

For a brief comparison of distribution angles with observations, we make use of projected misalignment angles as viewed from the $z$-axis ($\psi_{\rm{2D}}$), within a projected half-mass radii ($r_{50,z}$) in the $x-y$ plane, though our choice of axis is arbitrary as galaxies within \textsc{eagle} have no preferential orientation. 

\begin{figure}
    \begin{center}
    \includegraphics[width=8.5cm]{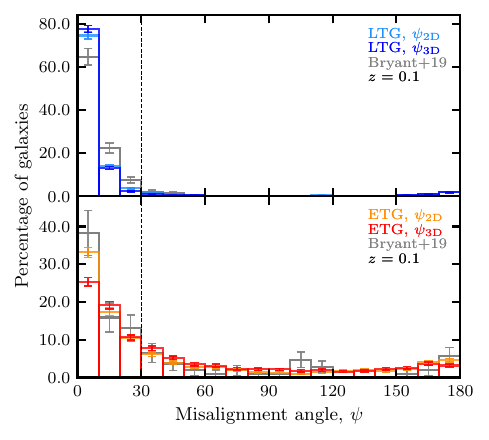}\\
    \caption{Projected 2D (cyan/orange) and 3D (blue/red) stellar-\gassf misalignment angle distributions for LTGs (top) and ETGs (bottom) for our sample of misalignments at $z=0.1$. The vertical dashed black line denotes the boundary between aligned ($<30^{\circ}$) and misaligned galaxy populations ($>30^{\circ}$). Observational results from \citet{Bryant2019} are included in grey and should be compared against $\psi_{\rm{2D}}$ distributions. Errors are given as Poisson uncertainties. The simulated and observed galaxies exhibit fairly similar distributions.}
    \label{plot: misalignment distributions}
    \end{center}
\end{figure}

Figure~\ref{plot: misalignment distributions} shows the projected and 3D misalignment distributions for LTGs (top) and ETGs (bottom) at $z=0.1$, with uncertainties given as Poisson errors. We include recent observational results from \citet{Bryant2019} that were obtained using the SAMI galaxy survey \citep[][]{Croom12, Bryant15}. 

For values in projection, we find $38.9\pm5.3$~per~cent of ETGs and $7.9\pm1.7$~per~cent of LTGs display stellar-gas kinematic misalignments. For comparison, observational misalignment fractions by \citet{Bryant2019} are reported as $\approx33$~per~cent (ETGs) and $\approx5$~per~cent (LTGs), but with a stellar mass range covering a larger range of galaxy masses $10^{8} \lesssim M_{*} \lesssim 10^{12}$ M$_{\odot}$. Both the fractions of misalignments and the general shape of the misalignment distributions are in good agreement with recent observational results by \citet{Bryant2019} with a small discrepancy for LTGs within $\psi<30^{\circ}$. We find that \textsc{eagle} LTGs tend to be more strongly aligned within $\psi_{\rm{2D}}<10^\circ$, while producing fewer LTGs with $10^\circ<\psi_{\rm{2D}}<30^\circ$. This difference could be caused by a combination of the different morphological classifications used in observations, the different constraints on misalignment angle uncertainties used between our results and those of observations, and/or the different mass ranges considered. Our overall results are also in good agreement with a number of other observational studies \citep[e.g.][]{Davis2011, BarreraBallesteros2014, BarreraBallesteros2015, Chen2016, Jin2016, Ristea2022, Raimundo2023}, as well as with predictions from cosmological simulations \citep[e.g.][]{Khim2020, Casanueva2022}.

We find a small peak in the counter-rotating region for both morphologies, with projected counter-rotating fractions of $10.9\pm1.3$~per~cent and $3.3\pm0.5$~per~cent for ETGs and LTGs, respectively. These were not reported in previous results in \textsc{eagle} \citep[see][]{Casanueva2022}, or in the \textsc{horizon-agn} simulation \citep[see][]{Khim2020}. We visually inspected our sample of $\psi_{\rm{2D}}>170^{\circ}$ misalignments and confirmed these systems to be physical and misaligned in both 3D and projection. As a whole, we find counter-rotation present in $5.7\pm0.6$~per~cent of our sample. To test the significance of this result, we perform a Kolmogorov–Smirnov (KS) test in the range $[135^{\circ}-180^{\circ}]$ and compare this to a uniform distribution in the same range. We obtain KS-statistic $=0.19$ and p-value $=1.8\times10^{-9}$ for ETGs and KS-statistic $=0.41$ and p-value $=3.3\times10^{-18}$ for LTGs, indicating a statistically significant population of counter-rotators.

We find a 3D misalignment fraction of $44.9\pm5.4$~per~cent for ETGs and $7.2\pm0.5$~per~cent for LTGs. The overall shape of the $\psi_{\rm{3D}}$ misalignment distribution is similar to that of projected values, being almost identical for LTGs but with a smaller fraction of aligned ETGs in the range $[0^{\circ}-10^{\circ}]$. We also see a small decrease in the fraction of counter-rotating systems when measured in 3D, but the number of counter-rotating systems remains significant with respect to to a uniform distribution using a KS test. These results suggest that although bias from projected viewing angles can inflate the number of aligned ETGs, the use of projected angles does not affect the overall shape or significantly reduce the number of misaligned systems.

Notably absent from our results is a small peak in the polar ($\sim90^{\circ}$) regime, found in observational results by \citet{Bryant2019}. As explained in \citet{Khim2020}, this may be due to the resolution limit of $\sim1$~pkpc that is used for force calculations in many cosmological simulations. For galaxies with radii $\sim4-8$~pkpc, this leads to significantly thicker overall discs in relation to their extent \citep[known as the `thin disc' problem in simulations; e.g.][]{Trayford2017}. For galaxies that are residing in the polar regime, their relative stability comes from the symmetrical torques experienced by their thin discs. However, with a disc thickness of $\sim1$ pkpc, it may become harder to achieve this symmetry and thus meta-stability, with these systems therefore decaying faster when compared to real galaxies. 

\begin{figure}
    \begin{center}
    \includegraphics[width=8.5cm]{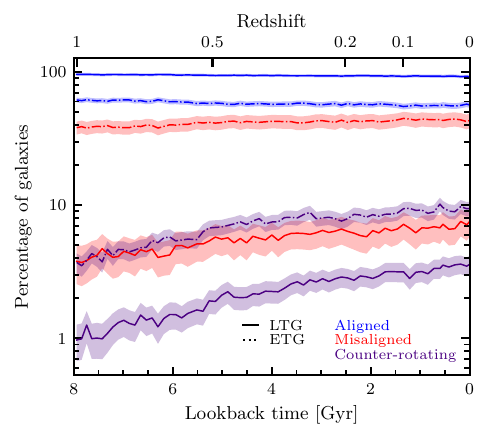}\\
    \caption{Stellar-\gassf misalignment angle fractions for LTGs (solid lines) and ETGs (dash-dotted lines) between $0<z<1$. Percentages are given with respect to the total galaxies of a given morphology at a given redshift. We classify galaxies as aligned with $\psi_{\rm{3D}}<30^{\circ}$ (blue), misaligned galaxies with $\psi_{\rm{3D}}>30^{\circ}$ (red), and counter-rotating galaxies with $\psi_{\rm{3D}}>150^{\circ}$ (purple). Errors are given as Poisson uncertainties (shaded regions). We see a slight upward trend in the misalignment fraction for both ETGs and LTGs across this redshift range, coinciding with an increase in counter-rotating fraction for both galaxy populations.}
    \label{plot: misalignment distribution evolution}
    \end{center}
\end{figure}

To justify our decision to consider both the co-rotating and counter-rotating regimes as stable, we show in Figure~\ref{plot: misalignment distribution evolution} how the fraction of aligned, misaligned, and counter-rotating systems varies for all snipshots between $0<z<1$. We consider the morphology independently and percentages are given with respect to the total number of galaxies of that morphological sub-sample. 

For both ETGs and LTGs we see a slight upward trend in the misalignment fraction toward $z=0$. This increase appears to result overwhelmingly from the build-up of long-lived counter-rotating systems. This is seen most clearly for ETGs, in which the $\approx5.1$~per~cent increase in misalignment fraction is captured by an $\approx5.7$~per~cent increase in counter-rotating fraction. The build-up of these systems appears to indicate that counter-rotating systems are in fact long-lasting and dynamically stable, in agreement with theoretical predictions \citep[][]{Tohline1982, LakeNorman1983}. We thus decide to treat these as a separate stable regime among the misaligned galaxy population. Some fraction of these systems may return to a co-rotating state upon another stochastic gas accretion event or undergo an apparent stellar angular momentum flip as the gas$_{\mathrm{SF}}$ forms a counter-rotating stellar disc that becomes dominant. For this latter process to occur, however, the masses of the counter-rotating gas and the stars have to be similar and the system must remain counter-rotating for long enough to form a dominant population of counter-rotating stars \citep[for an observational counterpart see e.g. NGC 4550; ][]{Rubin92, Rix92, Wiklind01}. We note that we see these apparent stellar angular momentum flips for several of our galaxies, as shown and discussed in Section~\ref{results: galaxy evolutions}. 

\begin{figure*}
    \hspace*{0cm}
    \includegraphics[width=8.7cm]{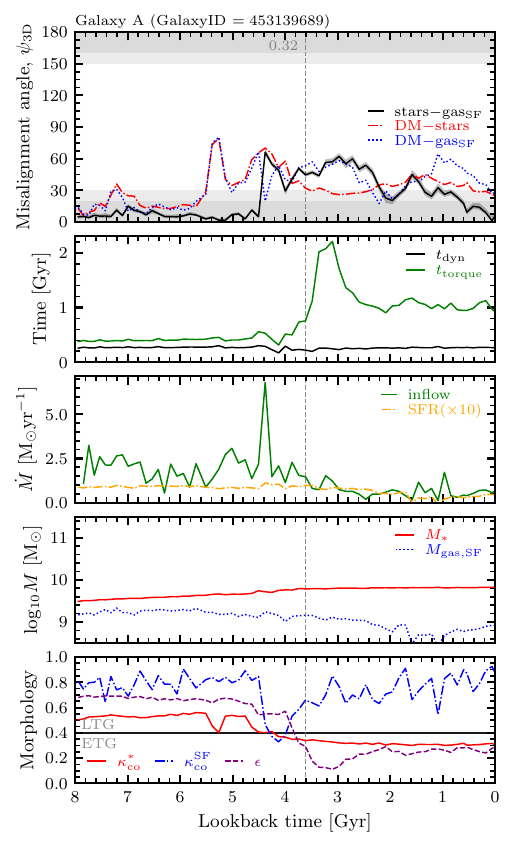} 
    \includegraphics[width=8.7cm]{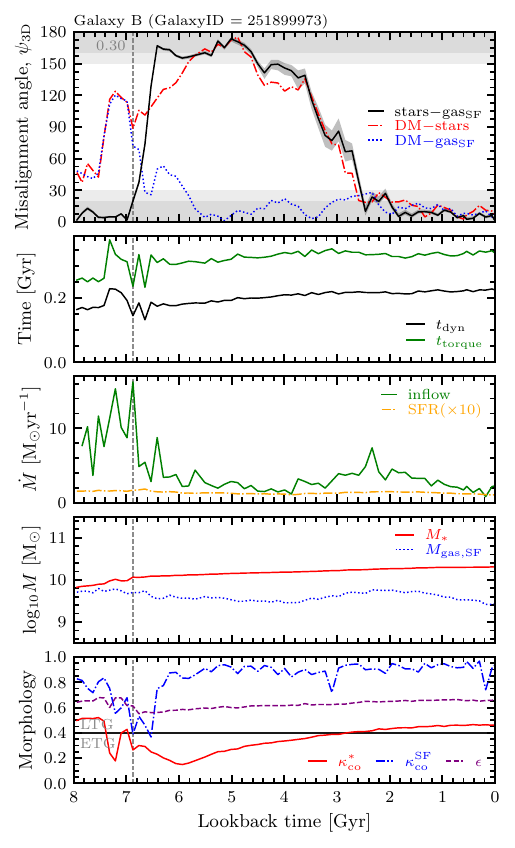}
    \\
    \caption{Evolution of two galaxies that experience long relaxation periods. Left: Galaxy A displaying an unstable misalignment matching our criteria with $t_{\rm{relax}}\approx3.78$~Gyr $\approx15.0\bar{t}_{\rm{dyn}}\approx3.5\bar{t}_{\rm{torque}}$. Right: Galaxy B shows an initial unstable misalignment and relaxation into the stable counter-rotating regime with $t_{\rm{relax}}\approx0.35$~Gyr $\approx2.0\bar{t}_{\rm{dyn}}\approx1.1\bar{t}_{\rm{torque}}$, before the cumulative effects of star formation induce a second unstable misalignment driven by a stellar angular momentum flip back into a co-rotating state with $t_{\rm{relax}}\approx2.55$~Gyr $\approx12.0\bar{t}_{\rm{dyn}}\approx7.5\bar{t}_{\rm{torque}}$. In order, the rows show: (1) the 3D misalignment angle between stars and \gassf (black) within $r_{50}$ ($1\sigma$ error bars shown in grey) and the 3D misalignment angle between the stars/\gassf at $r_{50}$ and the DM at 30 pkpc, with the horizontal dark and light grey shaded regions representing the stable regime ($<20^{\circ}$ and $>160^{\circ}$) and aligned/counter-rotating regime ($<30^{\circ}$ and $>150^{\circ}$), respectively}, (2) the dynamical time (solid green line) and torquing time (solid black line) of the system, (3) the total gas inflow rate at $r_{50}$ (solid green line)  and the SFR multiplied by a factor of 10 (dashed orange line) within $2r_{50}$, (4) the stellar mass (solid red line) and \gassf mass (dotted blue line) within $2r_{50}$, (5) the stellar co-rotational energy (red solid line) that we use as a proxy for morphology (with the horizontal black line marking the boundary between ETGs and LTGs), the \gassf co-rotational energy within $r_{50}$ (blue dash-dot line), and the stellar ellipticity (purple dashed line). Mergers are indicated by the vertical grey dashed line and occur at $t_{\rm{lookback}}=3.61$ ($\mu_{\rm{*}}=0.32$) for Galaxy A, and $t_{\rm{lookback}}=6.88$ ($\mu_{\rm{*}}=0.30$) for Galaxy B.
    \label{plot: evolutions}
\end{figure*}

The longevity of counter-rotating systems in \textsc{eagle} is also in agreement with results obtained using other cosmological hydrodynamical simulations \citep[e.g.][]{Starkenburg2019, Cenci2023}. \citet{Starkenburg2019} use the IllustrisTNG \citep{Nelson19} hydrodynamical simulations to investigate the longevity of counter-rotating systems, finding these to be long-lasting relative to other orientations. Similarly, \citet{Cenci2023} investigated the formation of a long-lived counter-rotating galaxy within the FIREbox simulations \citep[][]{Feldmann2023}. Following a starburst induced by the interaction with a nearby low-mass satellite and the subsequent rapid depletion of co-rotating gas, a long-lived ($\sim4$~Gyr) counter-rotating gas disc is able to form from the accretion of stripped misaligned gas. We note that carrying out a detailed analysis of this behaviour in \textsc{eagle} is beyond the scope of this work.

\subsection{Formation and relaxation process of stellar-gas kinematic misalignments}\label{results: galaxy evolutions}

\begin{figure*}
    \hspace*{0cm}
    \includegraphics[width=\textwidth]{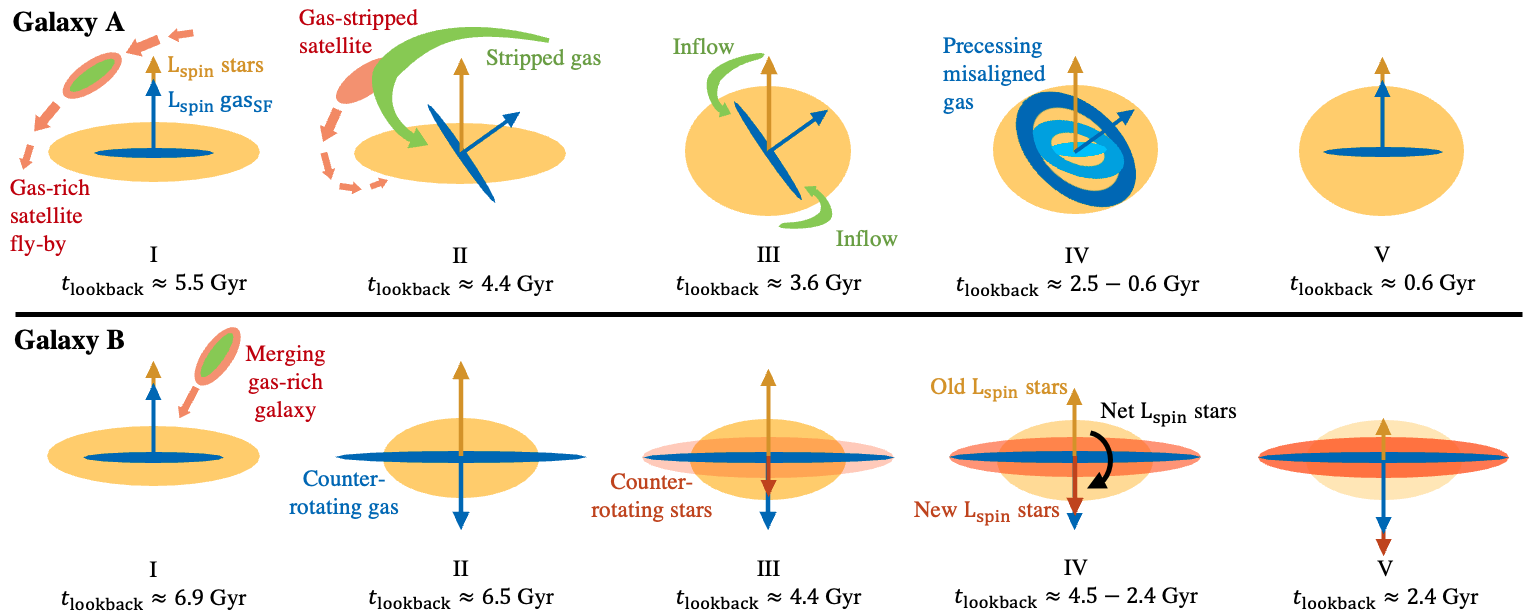} 
    \\
    \caption{Illustration of the evolution of two galaxies shown in Figure~\ref{plot: evolutions} in five key stages. The net angular momentum vectors of the stellar (yellow) and \gassf (blue) are represented by arrows. Top: evolution of Galaxy A beginning with (I) an initial flyby of a gas-rich satellite that causing a stellar-DM misalignment; (II) the formation of an unstable misalignment from the accretion of gas stripped from the satellite during a second fly-by; (III) merging of the satellite galaxy with the central galaxy and the resulting morphological transformation of the galaxy to a more spheroidal stellar mass distribution, while maintaining the unstable misalignment through continued accretion of stripped gas; (IV) a decrease in the inflow rate allows the gas disc to begin precessing and relaxing into the galactic plane; (V) the gas disc becomes aligned with the stellar angular momentum. Bottom: evolution of Galaxy B beginning with (I) a gas-rich merger that disrupts the existing gas disc; (II) the quick formation of a large misaligned gas disc that quickly settles to become counter-rotating; (III) a new generation of counter-rotating stars begin to form from the counter-rotating gas which begins to build a new counter-rotating stellar disc; (IV) the stellar angular momentum eventually flips, manifesting as a `pseudo-relaxation'; (V) the counter-rotating stars dominate the net angular momentum and the galaxy appears as aligned.}
    \label{plot: galaxy A B graphic}
\end{figure*}

To highlight an example of a long-lived unstable misalignment and to illustrate our method involving the identification and extraction of unstable misalignment timescales, we provide examples of two galaxy evolutions in Figure~\ref{plot: evolutions}. Illustrations of these evolutions are provided in Figure~\ref{plot: galaxy A B graphic} showing five key stages in each galaxy's evolution (labelled \RNum{1}-\RNum{5}). We will refer to these stages throughout this subsection. Both galaxies experience at least one unstable misalignment with a relaxation time $t_{\rm{relax}}>2$~Gyr. These galaxies were selected for their interesting stellar-\gassf kinematic evolution and the longevity of their unstable misalignments (as detailed below). We stress that the exceptionally long relaxation times of the two selected galaxies tend to be a rarity among our sample, as we discuss in detail in Section~\ref{results: relaxation timescales}.

Galaxy A (left panels of Figure~\ref{plot: evolutions}) is a relatively low-mass galaxy with $M_{*}\approx8\times10^{9}$~M$_{\odot}$ at $z=0$ and with a single long-duration unstable misalignment starting at a lookback time $t_{\mathrm{lookback}}=4.38$~Gyr. (\RNum{1})~Around 1~Gyr before the unstable misalignment, this galaxy experiences a close encounter with a lower-mass ($M_{*}\approx3\times10^{9}$~M$_{\odot}$) satellite which decouples the angular momentum of the galaxy from its wider DM halo \citep[see also e.g.][]{Velliscig2015}. This is seen as a steep increase in the DM-stars and DM-\gassf misalignment angle illustrated in the top-left panel of Figure~\ref{plot: evolutions}. The galaxy remains marginally misaligned with its dark matter halo for the remainder of its evolution, with an average stellar-DM misalignment angle of $\approx30^{\circ}$. (\RNum{2})~A second close fly-by strips a large amount of gas from the satellite. Most of this gas accretes onto the central galaxy (seen as a sharp increase in gas inflow rate), disrupting the existing co-rotating gas disc (seen as a decrease in $\kappa_{\rm{co}}^{\rm{SF}}$), and quickly forms an unstable misalignment. (\RNum{3})~Ongoing accretion of stripped gas immediately after the misalignment formation is able to sustain the angular momentum of the misaligned gas disc, maintaining a steady misalignment angle of $\psi_{\rm{3D}}\approx60^{\circ}$ until $t_{\mathrm{lookback}}\sim2.5$~Gyr. (\RNum{4})~This is followed by a decrease in the gas inflow rate and the onset of differential precession about the stellar angular momentum axis. The unstable misalignment begins to relax toward the galactic plane, (\RNum{5})~eventually settling into the stable co-rotating regime at $t_{\mathrm{lookback}}=0.60$~Gyr with a $t_{\rm{relax}}\approx3.78$~Gyr. Expressed in terms of time-averaged \tdyn and \ttorque we find $t_{\rm{relax}}\approx15.0\bar{t}_{\rm{dyn}}\approx3.5\bar{t}_{\rm{torque}}$, indicating this unstable misalignment has persisted for significantly longer than predicted by a purely dynamical settling time. We note that the behaviour of this misalignment post-formation is similar to that studied by \citet{VoortDavis2015} using the \textsc{fire} \citep[][]{Hopkins14} cosmological zoom-in simulations. 

While in the process of relaxing, the galaxy undergoes a major merger with the previously gas-stripped satellite at $t_{\mathrm{lookback}}=3.61$~Gyr (horizontal dashed line in the left panels of Figure~\ref{plot: evolutions}) with a stellar mass ratio of $\mu_{\rm{*}}=0.32$. While this merger does not substantially change $t_{\rm{dyn}}$, it causes a significant morphological transformation from an LTG to an ETG (seen as a decrease in $\kappa_{\rm{co}}^*$; bottom-left panel of Figure~\ref{plot: evolutions}) and a decrease in the stellar ellipticity (see panel \RNum{3} in Figure~\ref{plot: galaxy A B graphic}). This drives a significant and permanent increase in $t_{\rm{torque}}$ and helps drive an overall longer relaxation time. This behaviour is not uncommon among other long-duration unstable misalignments. We find a $>20$~per~cent decrease in stellar ellipticity between pre- and post-unstable misalignment formation is seen in $\approx33$~per~cent of $t_{\rm{relax}}>2$~Gyr relaxations. Despite the increase in $\bar{t}_{\rm{torque}}$ to $1-2$~Gyr, the galaxy remains unable to fully relax for another $\approx3$~Gyr. This is due to a small gas accretion event at $t_{\mathrm{lookback}}\sim1.5$~Gyr, although the accretion rate is insufficient to prevent further differential precession and eventual relaxation into stable co-rotation.

Galaxy B (right panels of Figure~\ref{plot: evolutions}) shows a very different overall kinematic evolution. (\RNum{1})~Analogous to Galaxy A, this galaxy becomes misaligned with respect to its DM halo shortly before a gas-rich major merger ($\mu_{\rm{*}}=0.30$) at $t_{\mathrm{lookback}}=6.88$~Gyr. (\RNum{2})~This merger feeds an inflow of gas to the inner $r_{50}$, disrupting the existing \gassf and causing it to settle quickly into a retrograde disc in the short timescale of $t_{\rm{relax}}\approx0.35$~Gyr $\approx2.0\bar{t}_{\rm{dyn}}\approx1.1\bar{t}_{\rm{torque}}$. (\RNum{3})~The galaxy remains in a counter-rotating configuration for $\sim2$~Gyr. During this time, the retrograde cold gas forms a new counter-rotating component of younger stars alongside the existing older co-rotating stars. This is seen from the steady decrease in \gassf mass and a corresponding increase in stellar mass, together with an increase in stellar co-rotational energy transforming the system back into a rotationally-supported LTG over $\approx4$~Gyr. (\RNum{4})~Starting at $t_{\mathrm{lookback}}\approx4.4$~Gyr, we begin to see an increase in the $1\sigma$ uncertainty in the stellar-\gassf misalignment angle. This marks the turning point at which the angular momentum of the two stellar components compete over dominance, beginning the second unstable misalignment and realignment into a co-rotational state. As the apparent relaxation was driven by a change in the stellar net-angular momentum rather than an explicit relaxation of the gas disc, this event can be considered a 'pseudo-relaxation'. This stellar net-angular momentum flip (traced both by the stellar-\gassf angle and stellar-DM angle; see the top-right panel of Figure~\ref{plot: evolutions}) occurs gradually, as reflected by the long relaxation time $t_{\rm{relax}}\approx2.55$~Gyr $\approx12.0\bar{t}_{\rm{dyn}}\approx7.5\bar{t}_{\rm{torque}}$, starting at $t_{\mathrm{lookback}}=4.38$~Gyr. (\RNum{5})~By $t_{\mathrm{lookback}}\approx2.4$~Gyr the galaxy shows a clear alignment between the (now discy) stellar and \gassf components.

\begin{table}
 \caption{Statistics on the median and standard deviation (SD) of relaxation timescales for our sample of relaxation paths, expressed for the total sample, by morphology (e.g. ETG $\rightarrow$ ETG) and by relaxation path (e.g. co $\rightarrow$ counter denoting a relaxation from a co-rotating to counter-rotating state). Relaxation timescales are given in terms of~Gyr, and as normalised relaxation times in terms of $\bar{t}_{\rm{dyn}}$, and $\bar{t}_{\rm{torque}}$.}
     \label{tab: median relaxation times}
 \begin{tabular}{p{2.1cm}>{\centering}p{0.65cm}>{\centering}p{0.5cm}>{\centering}p{0.65cm}>{\centering}p{0.5cm}>{\centering}p{0.65cm}>{\centering\arraybackslash}p{0.5cm}}
 \toprule 
    \multirow{3}{*}{\centering Relaxation path} &
    \multicolumn{6}{c}{Relaxation time} \\
 \cmidrule(lr){2-7} &
    \multicolumn{2}{c}{All} &
    \multicolumn{2}{c}{LTG $\rightarrow$ LTG} &
    \multicolumn{2}{c}{ETG $\rightarrow$ ETG} \\
 \cmidrule(lr){2-3} \cmidrule(lr){4-5} \cmidrule(lr){6-7}
    &
    Median & SD &
    Median & SD &
    Median & SD \\
 \midrule
    $t_{\rm{relax}}$ (Gyr)        \\
    All                           & 0.50 & 0.64 &   0.38 & 0.36 &    0.51 & 0.68 \\
    co $\rightarrow$ co           & 0.47 & 0.66 &   0.31 & 0.34 &    0.51 & 0.68 \\
    counter $\rightarrow$ counter & 0.43 & 0.38 &   0.36 & 0.22 &    0.49 & 0.40 \\
    co $\rightarrow$ counter      & 0.51 & 0.57 &   0.43 & 0.39 &    0.68 & 0.65 \\
    counter $\rightarrow$ co      & 1.03 & 0.86 &   0.40 & 0.40 &    1.15 & 0.85 \\
 \midrule
    $t_{\rm{relax}}/\bar{t}_{\rm{dyn}}$ \\
    All                           & 3.15 & 5.07 &   2.50 & 3.47 &    3.34 & 5.29 \\
    co $\rightarrow$ co           & 2.75 & 5.11 &   1.87 & 3.53 &    3.06 & 5.22 \\
    counter $\rightarrow$ counter & 3.11 & 2.93 &   2.56 & 1.72 &    3.11 & 3.19 \\
    co $\rightarrow$ counter      & 4.45 & 4.48 &   3.66 & 3.24 &    5.33 & 4.74 \\
    counter $\rightarrow$ co      & 9.32 & 7.54 &   3.76 & 3.93 &    10.5 & 7.85 \\
 \midrule
    $t_{\rm{relax}}/\bar{t}_{\rm{torque}}$ \\
    All                           & 1.41 & 2.09 &   1.38 & 1.95 &    1.36 & 2.01 \\
    co $\rightarrow$ co           & 1.21 & 1.94 &   1.01 & 1.87 &    1.22 & 1.84 \\
    counter $\rightarrow$ counter & 1.45 & 1.45 &   1.27 & 1.03 &    1.44 & 1.56 \\
    co $\rightarrow$ counter      & 2.41 & 2.09 &   2.14 & 1.94 &    2.59 & 1.70 \\
    counter $\rightarrow$ co      & 4.69 & 3.66 &   2.43 & 2.43 &    4.86 & 3.80 \\
 \bottomrule
 \end{tabular}
\end{table}

\begin{figure}
    \begin{center}
    \includegraphics[width=8.5cm]{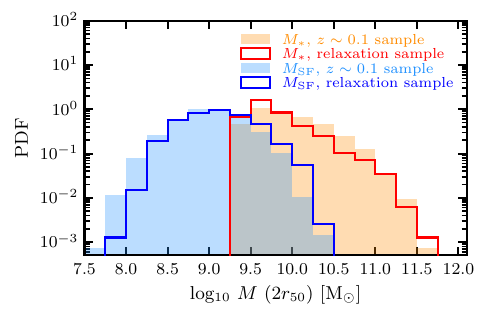}\\
    \caption{Stellar and \gassf mass distributions within $2r_{50}$ of our $z = 0.1$ unstable misalignment distribution sample (orange/light blue shaded regions) and our relaxation sample (red/blue lines) between $0<z<1$. We find our relaxation sample is broadly representative of the sample at $z = 0.1$ in terms of stellar mass, with a slight bias toward higher \gassf masses.}
    \label{plot: sample_vs_dist}
    \end{center}
\end{figure}

While both galaxies undergo mergers that coincide with their unstable misalignments, we only attribute a merger origin explicitly to the first unstable misalignment in Galaxy B. Both galaxies also show a prior DM halo decoupling preceding the formation of the misalignment, owing to previous galaxy interactions. As discussed later in Section~\ref{results: misalignment origins}, we find that mergers tend to be sub-dominant in the formation of misalignments among our sample. 

\subsection{Relaxation timescale distributions}\label{results: relaxation timescales}

Using our selection criteria (see Section~\ref{method: relaxation time definition}) we extract a sample of 3154 unstable misalignment relaxations between $0<z<1$. We include each instance of an unstable misalignment separately, even if these occur in the same galaxy (such as Galaxy B described in Section~\ref{results: galaxy evolutions}). Median values and statistics are found in Table~\ref{tab: median relaxation times} and Table~\ref{tab: relaxation sample}.

As illustrated in Figure~\ref{plot: sample_vs_dist}, this sample has a stellar and \gassf mass broadly representative of the sample used at $z=0.1$, with a small bias towards higher \gassf masses and thus \gassf fractions. This is in part due to the earlier redshifts covered, and partially as a consequence of requiring consecutive snipshots meeting our sample criteria for the duration of the unstable misalignment. In the case of the latter, this removes many galaxies that briefly fall below the minimum \gassf particle count whilst unstably misaligned. 

\subsubsection{The total galaxy population}\label{results: relaxation timescales total population}

\begin{figure}
    \begin{center}
    \includegraphics[width=8.5cm]{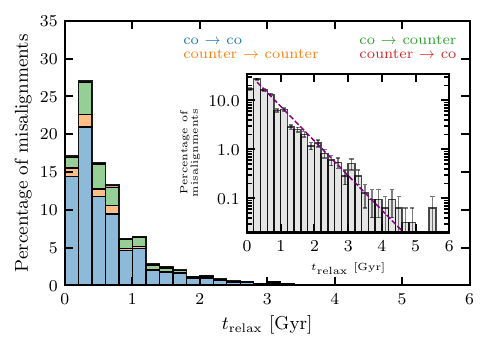}\\
    \includegraphics[width=8.5cm]{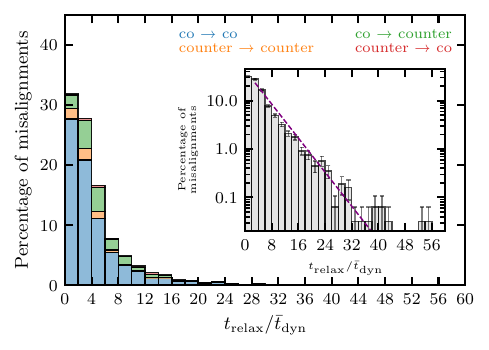}\\
    \includegraphics[width=8.5cm]{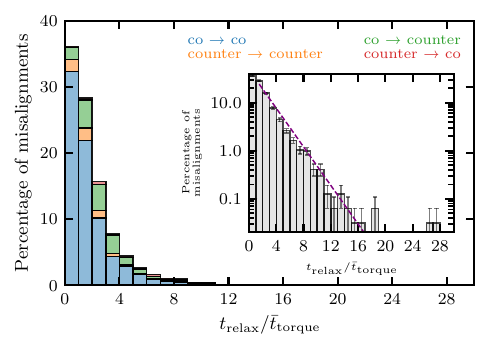}\\
    \caption{Distributions of relaxation times of our total sample of 3154 unstable misalignment relaxations for $0<z<1$. Relaxations are coloured according to the relaxation path: co-rotating to co-rotating (blue), co-rotating to counter-rotating (green), counter-rotating to counter-rotating (orange), and counter-rotating to co-rotating (red). Inset graphs show the same total distribution on a log percentage axis, with a line of best fit (dashed purple). Errors are given as Poisson uncertainties. We find a broadly log-linear distribution (inset graphs) of relaxation time distributions, with the majority of unstable misalignments relaxing within shorter timescales but with a small population of longer-lived misalignments.}
    \label{plot: relaxation distributions total}
    \end{center}
\end{figure}

Galaxies in \textsc{eagle} display a variety of relaxation times and pathways. Figure~\ref{plot: relaxation distributions total} shows the distributions of relaxation times (top), and normalised relaxation times in terms of $t_{\rm{relax}}/\bar{t}_{\rm{tdyn}}$ (middle), and as $t_{\rm{relax}}/\bar{t}_{\rm{torque}}$ (bottom) for our total sample of unstable misalignment relaxations. The majority of unstable misalignments in our sample tend to be fairly short-lived, with median relaxation times of $t_{\rm{relax}}=0.50$~Gyr $=3.2\bar{t}_{\rm{dyn}}=1.4\bar{t}_{\rm{torque}}$. 

Due to time resolution limitations, we naturally underestimate the number of very short-duration unstable misalignments that occur over less than 2 successive snipshots ($t_{\rm{relax}}\lesssim0.12$~Gyr), as seen from the first bin of the top plot in Figure~\ref{plot: relaxation distributions total}. Ignoring this bin, we find unstable misalignment distributions to be described by an approximately log-linear distribution of relaxation timescales (inset figures in Figure~\ref{plot: relaxation distributions total} with best-fit values given in Table~\ref{tab: best-fit lines}). These results (especially the shape of the $t_{\rm{relax}}/\bar{t}_{\rm{torque}}$ distribution) suggest that while some galaxies appear to experience unstable misalignments lasting longer than those predicted by theory \citep{Tohline1982, LakeNorman1983}, the majority of unstable misalignments are associated with relaxation times governed by a fast settling time in the absence of ongoing smooth accretion.

Focusing on the tail-end of the distributions, we find a moderate population ($\approx20.1$~per~cent) of unstable misalignments persisting for $t_{\rm{relax}}>1$~Gyr, with only a handful of examples exceeding $\gtrsim20\bar{t}_{\rm{dyn}}$. Additionally, we find $159$ ($5.0$~per~cent) instances of unstable misalignments with $t_{\rm{relax}}>2$~Gyr (two of which are shown and discussed in Section~\ref{results: galaxy evolutions}) that are approximately of similar duration to the galaxy investigated by \citet{VoortDavis2015} in the \textsc{fire} simulations \citep{Hopkins14}. The population of long-duration unstable misalignments becomes a little clearer when expressed as $t_{\rm{relax}}/\bar{t}_{\rm{torque}}$, with $\approx19.9$~per~cent of relaxations lasting more than three times longer ($>3\bar{t}_{\rm{torque}}$) than expected from a purely dynamical relaxation (see Section~\ref{theory:}). We note that while our relaxation selection method identifies two unstable misalignments with exceptionally enhanced relaxation timescales of $\gtrsim 
25t_{\rm{relax}}/\bar{t}_{\rm{torque}}$, closer inspection reveals these to be systems that spent considerable time ($>1.5$~Gyr) near the stable counter-rotating regime without settling. We attribute these outliers to the consequences of choosing a non-zero latency time.

\begin{table}
 \caption{Statistics on the numbers (N) and fractions ($\rm{f}_{\rm{sub}}$) of unstable misalignments in our relaxation sample. Statistics are subdivided by morphology (e.g.\ ETG~$\rightarrow$~ETG) and by relaxation path (e.g.\ co~$\rightarrow$~counter denoting a relaxation from a co-rotating to counter-rotating state). Fractions are given with respect to the total of a given morphology sub-sample.}
     \label{tab: relaxation sample}
 \begin{tabular}{lcccccc}
 \toprule 
    \multirow{2}{*}{Relaxation type} &
    \multicolumn{2}{c}{All} &
    \multicolumn{2}{c}{LTG $\rightarrow$ LTG} &
    \multicolumn{2}{c}{ETG $\rightarrow$ ETG} \\
 \cmidrule(lr){2-3} \cmidrule(lr){4-5} \cmidrule(lr){6-7}
    &
    N & $\rm{f}_{\rm{sub}}$ &
    N & $\rm{f}_{\rm{sub}}$ &
    N & $\rm{f}_{\rm{sub}}$ \\
 \midrule
    All                           & 3154 & 1.00 &  802 & 1.00 & 2007 & 1.00 \\
    co $\rightarrow$ co           & 2403 & 0.76 &  501 & 0.62 & 1669 & 0.83 \\
    counter $\rightarrow$ counter & 181  & 0.06 &   30 & 0.04 &  137 & 0.07 \\
    co $\rightarrow$ counter      & 499  & 0.16 &  262 & 0.33 &  146 & 0.07 \\
    counter $\rightarrow$ co      & 71   & 0.02 &    9 & 0.01 &   55 & 0.03 \\
 \bottomrule
 \end{tabular}
\end{table}

\begin{table*}
 \caption{Log-linear best-fit relationships expressed as a fraction of the respective relaxation (sub-)sample ($f_{\rm{sub}}$) corresponding to the total relaxation sample (first row) in Figure~\ref{plot: relaxation distributions total} and for sub-samples of LTG~$\rightarrow$~LTG and ETG~$\rightarrow$~ETG relaxations (rows 2 and 3) in Figure~\ref{plot: LTG ETG relaxation distributions}. Relaxation timescales are given in terms of~Gyr, and as normalised relaxation times in terms of $\bar{t}_{\rm{dyn}}$, and $\bar{t}_{\rm{torque}}$.}
     \label{tab: best-fit lines}
 \begin{tabular}{lccc}
 \toprule 
    \multirow{2}{*}{\centering Relaxation (sub-)sample} &
    \multicolumn{3}{c}{$\rm{log}_{10}(f_{\rm{sub}})$} \\
  \cmidrule(lr){2-2} \cmidrule(lr){3-3} \cmidrule(lr){4-4}
    &
    $t_{\rm{relax}}$ (Gyr) &
    $t_{\rm{relax}}/\bar{t}_{\rm{dyn}}$ &
    $t_{\rm{relax}}/\bar{t}_{\rm{torque}}$ \\
  \midrule
    All                         & $-0.71t_{\rm{relax}} - 0.42$ 
                                & $-0.09\bar{t}_{\rm{dyn}} - 0.37$ 
                                & $-0.20\bar{t}_{\rm{torque}} - 0.31$ \\
    LTG $\rightarrow$ LTG       & $-1.16t_{\rm{relax}} - 0.15$ 
                                & $-0.10\bar{t}_{\rm{dyn}} - 0.33$ 
                                & $-0.18\bar{t}_{\rm{torque}} - 0.36$ \\
    ETG $\rightarrow$ ETG       & $-0.66t_{\rm{relax}} - 0.45$ 
                                & $-0.08\bar{t}_{\rm{dyn}} - 0.38$ 
                                & $-0.21\bar{t}_{\rm{torque}} - 0.30$ \\
 \bottomrule
 \end{tabular}
\end{table*}

\subsubsection{Relaxation timescales with relaxation paths}\label{results: relaxation timescales with path}

The large majority of unstable misalignments return to their original stability regime with \coco and \countercounter relaxation paths making up $\approx76$~per~cent and $\approx4$~per~cent of our sample respectively (see Table~\ref{tab: relaxation sample}). Co $\rightarrow$ counter relaxations make up $\approx15.8$~per~cent, while we find only 71 instances ($\approx2.3$~per~cent) of \counterco relaxations. Given that only $\approx6$~per~cent of galaxies in our sample at a given snipshot are counter-rotating, it is unsurprising that we find significantly fewer unstable misalignments originating from the counter-rotating stable regime. 

\begin{figure*}
    \hspace*{0cm}
    \includegraphics[width=17.8cm]{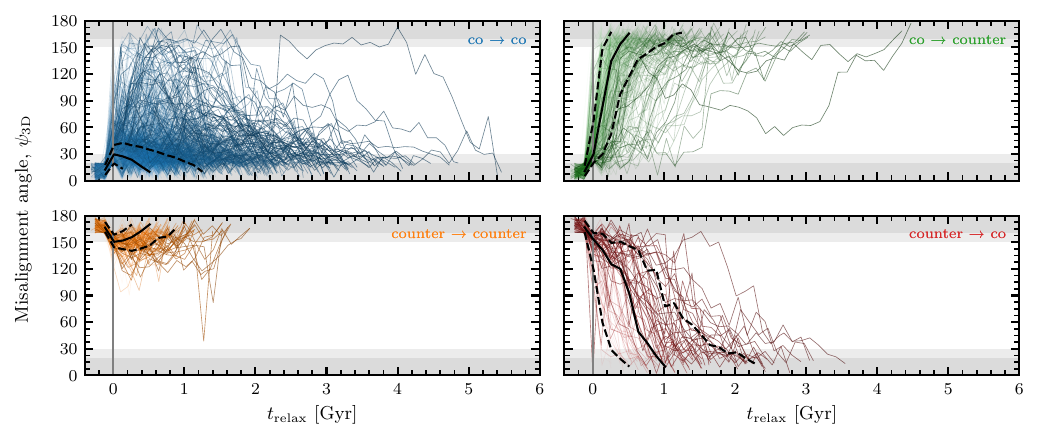}\\
    \caption{Unstable misalignment relaxation pathways between $0<z<1$, showing co-rotating to co-rotating relaxations (upper left), co-rotating to counter-rotating relaxations (upper right), counter-rotating to counter-rotating relaxations (lower left), and counter-rotating to co-rotating relaxations (lower right). Relaxation paths are stacked at the first snipshot in the unstable regime $[20^{\circ} - 160^{\circ}]$ (vertical grey line). Unstable misalignments will eventually relax back into the aligned ($<30^{\circ}$) or counter-rotating ($>150^{\circ}$) regimes (light grey shaded regions). Galaxies are assumed to have relaxed upon spending $>0.1$~Gyr in a consecutive stable regime of $<20^{\circ}$ or $>160^{\circ}$ (dark grey shaded regions). Median relaxation paths (solid black line) and $1\sigma$ percentile relaxation paths (dashed black lines) are shown for each relaxation type. Longer relaxation times have had their hue darkened to stand out, while short relaxation times have increased transparency.}
    \label{plot: stacked all}
\end{figure*}

In Figure~\ref{plot: stacked all} we visualise the relaxation paths for our total relaxation sample alongside their respective median pathways. We find the relaxation time to vary depending on the relaxation path taken (Table~\ref{tab: median relaxation times}). Unstable misalignments that returned to their original stability regime (\coco and counter $\rightarrow$ counter) tend to relax faster while unstable misalignments that end up transitioning between stability regimes (\cocounter and counter $\rightarrow$ co) tend to be longer in duration. This trend is also seen for normalised relaxation timescales. Expressed in terms of $\bar{t}_{\rm{torque}}$, we find relatively short median relaxation timescales for \coco ($1.21\bar{t}_{\rm{torque}}$) and \countercounter ($1.45\bar{t}_{\rm{torque}}$) relaxation paths, and relatively long median relaxation timescales for \cocounter ($3.03\bar{t}_{\rm{torque}}$) and \counterco ($5.22\bar{t}_{\rm{torque}}$). 

\begin{figure}
    \begin{center}
    \includegraphics[width=8.5cm]{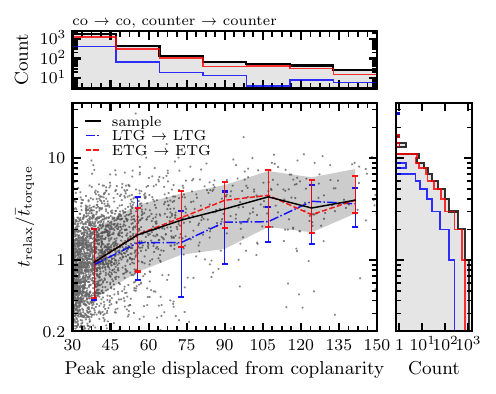}\\
    \caption{Peak misalignment angle from $0^{\circ}$ (co $\rightarrow$ co relaxation paths) and $180^{\circ}$ (counter $\rightarrow$ counter relaxation paths) reached during the relaxation process as a function of the normalised relaxation time for a sub-sample of \coco and \countercounter relaxation paths. Normalised median relaxation times are shown for our total sample (black solid line) with $1\sigma$ uncertainties (shaded grey), as well as exclusively LTGs (blue solid line) and ETGs (red solid line). Galaxies with unstable misalignments displaced further from coplanrity with the stellar disc are found to take longer to relax, plateauing once the polar-ring regime is reached. ETGs categorically have longer normalised relaxation times for greater displacements than LTGs up to $\approx90^{\circ}$ beyond which a low sample of LTGs limits the degree to which ETGs and LTGs can be distinguished.}
    \label{plot: peak displacement angle}
    \end{center}
\end{figure}

This result is not surprising given an unstable misaligned gas disc displaced by a larger angle from being coplanar with the stellar disc must dissipate more angular momentum to relax \citep[see Appendix 2 of ][]{Bryant2019}. We briefly highlight this behaviour using our sample of \coco and \countercounter relaxations in Figure~\ref{plot: peak displacement angle}, as these paths return to the original stability regime from which they were displaced. This allows us to characterise a maximum displaced angle from coplanarity with the stellar disc (i.e. $0^\circ$ for \coco and $180^\circ$ for \countercounter relaxations).

We see a clear positive relationship between the relaxation time and the displaced angle up to $\approx90^{\circ}$, showing that unstable misalignments displaced by a larger amount from the nearest stability regime will take longer to relax. We also see a population of unstable misalignments that return to their original stability regime despite being displaced angles of $>90^{\circ}$. To first order, unstable misalignments are expected to relax toward the nearest stable regime. This population of unstable misalignments displaced by $>90^{\circ}$ is, therefore, indicative of additional processes acting on the gas (e.g. smooth misaligned accretion) that prevent it from relaxing toward the nearest dynamically stable regime. Unstable misalignments displaying this behaviour tend to show significant enhancements to their relaxation timescales with $\gtrsim3\bar{t}_{\rm{torque}}$.

We see from Figure~\ref{plot: stacked all} that \coco and \countercounter relaxation paths show similar behaviour and the majority are displaced by only a small angle from the stable co- and counter-rotating regimes, respectively. As small displacements require less angular momentum dissipation, shorter relaxation timescales are expected for the average \coco and \countercounter relaxation paths in our sample. For instance, the large majority ($\approx73$~per~cent) of $t_{\rm{relax}}/\bar{t}_{\rm{torque}}<1$ relaxations are displaced by an angle of $<40^{\circ}$ from coplanarity.

For relaxations that transition \textit{between} stable regimes, this automatically involves displaced angles of $>90^{\circ}$. As more angular momentum must be dissipated in order to relax into one of the stable regimes, \cocounter and \counterco relaxation paths are expected to return longer relaxation timescales. We also see that \coco and \countercounter relaxations follow similar paths. This makes sense as \textsc{eagle} models stellar particles as non-collisional. Thus, to first order, co-rotating and counter-rotating systems can be considered dynamically-equivalent from the perspective of a misaligned gas particle in a polar orbit. 

Interestingly, we do not see the same similarity between \cocounter and \counterco relaxation paths, with the latter taking significantly longer to relax back into a co-rotating state. From visual inspection of our sample of 71 \counterco misalignments, we find that over half are likely driven primarily through changes in the net stellar angular momentum rather than explicit gas-disc relaxations (as found in Galaxy B in Figure~\ref{plot: evolutions}). These `apparent' relaxations would be expected to relax on timescales on the order of $\sim1$~Gyr, governed by the rate of conversion of counter-rotating star-forming gas into a dominant population of counter-rotating stars. This process is largely exclusive to \counterco relaxations given that it requires a substantial counter-rotating gas reservoir to begin with. 

\begin{figure*}
    \hspace*{0cm}
    \includegraphics[width=8.5cm]{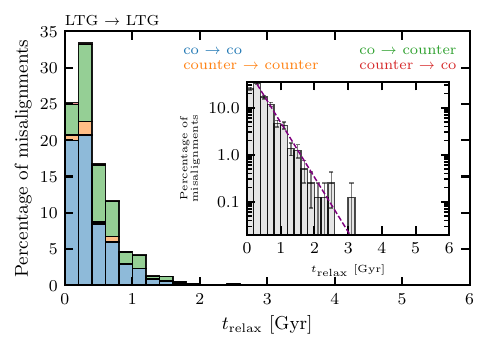}
    \includegraphics[width=8.5cm]{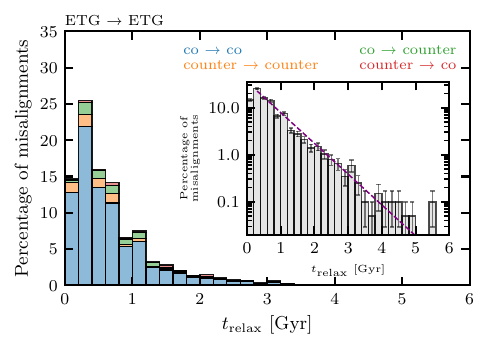}\\
    \includegraphics[width=8.5cm]{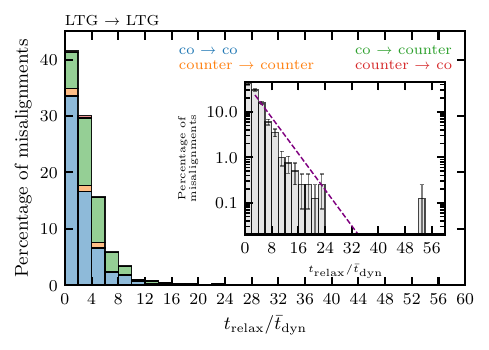}
    \includegraphics[width=8.5cm]{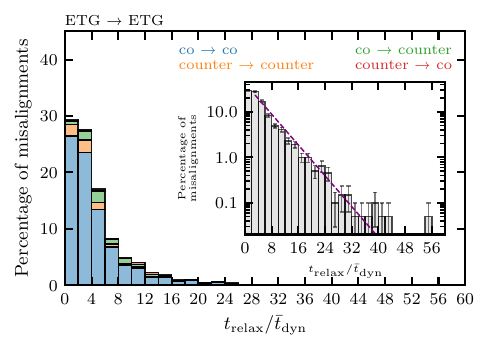}\\
    \includegraphics[width=8.5cm]{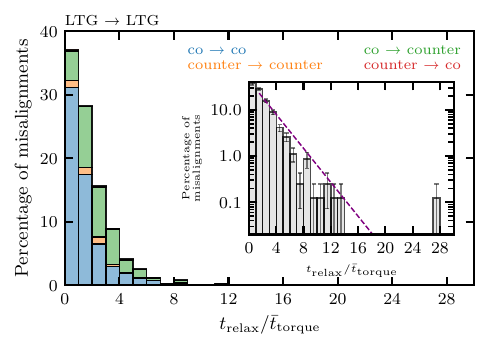}
    \includegraphics[width=8.5cm]{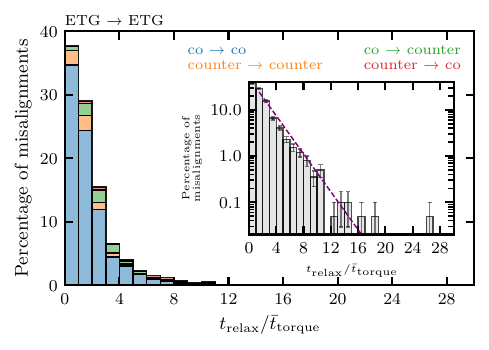}\\
    \caption{Distributions of relaxation times (top), and normalised relaxation times of $t_{\rm{relax}}/\bar{t}_{\rm{dyn}}$ (middle) and $t_{\rm{relax}}/\bar{t}_{\rm{torque}}$ (bottom). The difference in relaxation timescales and path composition for each galaxy morphology are shown for galaxies with \LTGLTG relaxations (left) and \ETGETG relaxations (right). Percentages are given with respect to the total number of relaxations within the morphological sub-sample. Relaxations are coloured according to the relaxation path, with co-rotating to co-rotating (blue), co-rotating to counter-rotating (green), counter-rotating to counter-rotating (orange), and counter-rotating to co-rotating (red). Inset graphs show the same distributions on a log percentage axis, with a line of best fit (dashed purple) showing a broadly log-linear relationship (values given in Table~\ref{tab: best-fit lines}). Errors are given as Poisson uncertainties.}
    \label{plot: LTG ETG relaxation distributions}
\end{figure*}

\subsubsection{Morphological dependence on relaxation timescales}\label{results: relaxation timescales with morphology}

We established in Section~\ref{results: misalignment distributions} that there is a significant morphological dependence on the fraction and shape of the misalignment distributions in \textsc{eagle}, echoing prior observational results \citep[e.g.][]{Bryant2019, Ristea2022, Casanueva2022}. We therefore also consider a possible morphological difference in relaxation time distributions and composition of relaxation paths between ETGs and LTGs. Henceforth we refer to ETG relaxations as those that explicitly begin and end their relaxation as ETGs (denoted as ETG $\rightarrow$ ETG), and similarly for LTGs (see Section~\ref{method: properties during misalignment} for details). We find a total of 802 LTG and 2007 ETG relaxations in our sample (see Table~\ref{tab: relaxation sample}), with the remaining 345 classified as having transitioned between morphologies after relaxation.

In Figure~\ref{plot: LTG ETG relaxation distributions}, we show the relaxation time and normalised relaxation time distributions split according to morphology. Median and standard deviations for these distributions are listed in Table~\ref{tab: median relaxation times}. We find some notable differences between LTG and ETG relaxation distributions, though both continue to be best described by log-linear distributions (see Table~\ref{tab: best-fit lines}). ETGs show on average longer relaxation times ($t_{\rm{relax}}=0.51$~Gyr) compared to LTGs ($t_{\rm{relax}}=0.38$~Gyr), with a shallower log-linear best-fit relationship. This distinction between ETGs and LTGs becomes weaker once the relaxation time is normalised in terms of \tdyn and $\bar{t}_{\rm{torque}}$, taking into account the impact of stellar morphology on the theoretical relaxation timescales.

To test for significance between these distributions, we run a two-sample KS test between LTG and ETG \trelax distributions. We find significant results for $t_{\rm{relax}}$ (KS-test statistic $=0.22$, p-value $=7.6\times10^{-25}$), with a decreasing significance for $t_{\rm{relax}}/\bar{t}_{\rm{dyn}}$ (KS-test statistic $=0.15$, p-value $=2.3\times10^{-11}$). We find no significant difference between morphologies in terms of $t_{\rm{relax}}/\bar{t}_{\rm{torque}}$ (KS-test statistic $=0.03$, p-value $=0.85$). This likely encapsulates the greater ability of $t_{\rm{relax}}/\bar{t}_{\rm{torque}}$ to account for dynamical and morphological properties of galaxies. We find no changes in the significance of the results using a stricter morphological classification of $\bar{\kappa}_{\rm{co}}^{\rm{*}}<0.35$ for ETGs and $\bar{\kappa}_{\rm{co}}^{\rm{*}}>0.45$ for LTGs.

\begin{figure}
    \begin{center}
    \includegraphics[width=8.5cm]{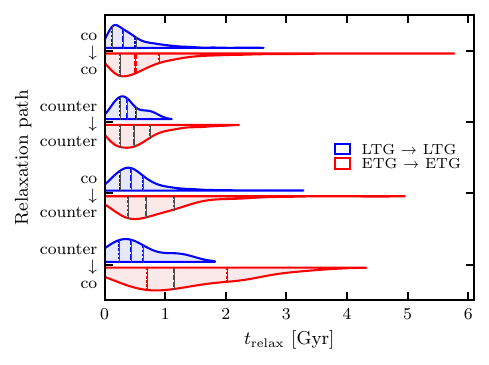}\\
    \includegraphics[width=8.5cm]{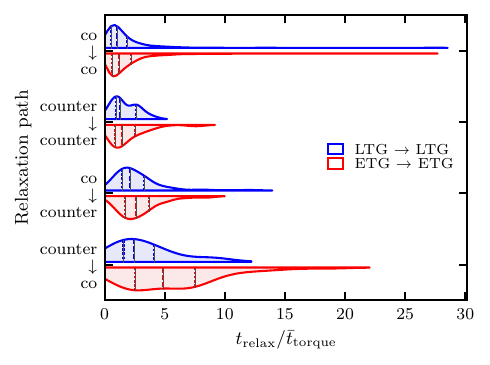}\\
    \caption{Relaxation timescale distributions (top) and normalised relaxation timescale distributions (bottom) for LTGs (blue) and ETGs (red) split by relaxation path. Statistics are visualized with dotted and dashed lines showing quartiles and the median respectively. Overall, \cocounter and \counterco relaxations tend to be longer in duration regardless of morphology. Accounting for different paths, ETGs tend to show longer relaxation times compared to LTGs. This difference is significantly weaker once morphological differences are considered using normalised relaxation times in terms of $\bar{t}_{\rm{torque}}$.}
    \label{plot: morph whisker plots}
    \end{center}
\end{figure}

Interestingly, while the shape of the $t_{\rm{relax}}/\bar{t}_{\rm{torque}}$ distributions and the gradient of the log-linear relationship describing the distribution is similar between morphologies, the composition of relaxation paths is different (as seen clearly in Figure~\ref{plot: LTG ETG relaxation distributions}, with compositional fractions in Table~\ref{tab: relaxation sample}). For instance, \cocounter relaxations constitute $\approx33$~per~cent of detected LTG misalignments, compared to only $\approx16$~per~cent in ETGs. The longer inherent relaxation times of \cocounter relaxation paths (as highlighted in Section~\ref{results: relaxation timescales with path}) dominate the tail-end of the distributions for LTGs. This tends to increase the overall median relaxation time of LTGs ($0.38$~Gyr).

At first glance, this path dichotomy between ETGs and LTGs appears unusual considering we also see $\sim1.8$ times the number (262) of \cocounter relaxations in LTGs compared to ETGs, but this is not seen for any other relaxation path (see Table~\ref{tab: relaxation sample}). This is despite our sample containing more than double the number of ETGs compared to LTGs. These results may suggest that LTGs preferentially transition between regimes, more so than ETGs. This seems unintuitive at first glance, given the typically larger gas discs in LTGs sets a larger angular momentum threshold needed to be overcome by accreting misaligned gas \citep{Khim2021}. This, in turn, should make these systems more resilient to larger misalignments that may cause the gas disc to become counter-rotating. 

A physical explanation may lie in the underlying differences in size and mass of the \gassf disc between morphologies. Given LTGs typically host larger star-forming gas discs, this sets a larger gas accretion threshold for the formation of unstable misalignments. This likely lowers the relative number of \coco misalignments that are otherwise common among ETGs. This could suggest larger stochastic accretion events such as gas-rich mergers or accretion through galaxy interactions are responsible for many of these misalignments in LTGs. 

Furthermore, \textsc{eagle} reproduces the observed galaxy morphology-density relation \citep{Dressler80}, with LTGs found to preferentially reside in lower density environments \citep{Pfeffer2023}. Compared to virialized high-density environments (which are dominated by gas-poor ETGs), galaxy interaction rates tend to be higher in lower-density environments \citep[][]{Ghigna1998, Pfeffer2023}. Likewise, the larger gas discs found in many LTGs create larger collisional cross-sections for gas stripping and mass exchange during close encounters with nearby gas-rich satellites. These encounters can significantly perturb the outskirts of the gas disc, which can manifest as an observable stellar-gas misalignment \citep[e.g.][]{Lu2021, Khim2021}. As gas accretion from galaxy encounters can be more stochastic and violent (compared to the smooth accretion of e.g. halo cooling), these encounters may have a greater likelihood to flip the star-forming gas disc into a counter-rotating state, which disproportionately affects LTGs. This may hint at an underlying difference in drivers of misalignments between morphologies depending on the large-scale environment. 

\subsubsection{Relaxation paths and morphology}\label{results: relax morphology and paths}

In Figure~\ref{plot: morph whisker plots} we compare the relaxation times of individual relaxation paths between morphologies. Considering relaxation paths separately, we see a trend toward longer median relaxation times in ETGs compared to LTGs (see Table~\ref{tab: median relaxation times}). For the dominant \coco relaxation paths in our sample, ETGs take $\approx1.5$ times longer to relax compared to LTGs. Testing for significance between morphologies in terms of $\bar{t}_{\rm{relax}}$ returns significant results for each path; \coco (KS-test statistic $=0.25$, p-value $=2.2\times10^{-21}$), \countercounter (KS-test statistic $=0.29$, p-value $=2.3\times10^{-2}$), \cocounter (KS-test statistic $=0.30$, p-value $=9.1\times10^{-8}$), and \counterco (KS-test statistic $=0.54$, p-value $=1.3\times10^{-2}$).  

While we found no significant difference between morphologies in the overall normalised relaxation time distributions in terms of $t_{\rm{relax}}/\bar{t}_{\rm{torque}}$ (see Section~\ref{results: relaxation timescales with morphology}), we do find a difference when \coco relaxation paths are considered separately. For \coco relaxation paths, ETGs experience $\approx1.2$ times longer normalised relaxation times compared to LTGs and this result is significant (KS-test statistic $=0.11$, p-value $=1.5\times10^{-4}$). Median \cocounter relaxation paths in ETGs are also $\approx1.2$ longer compared to LTGs in terms of $t_{\rm{relax}}/\bar{t}_{\rm{torque}}$. However, this result narrowly falls outside of being significant (KS-test statistic $=0.13$, p-value $=6.2\times10^{-2}$). The remaining relaxation paths do not show any differences between ETGs and LTGs that return significant KS-test results.

These results suggest that differences in overall median relaxation time values for ETGs and LTGs are driven in large part by morphological differences in the mass distributions of these systems. However, at least for \coco relaxation paths, we see a small preference toward processes preferentially preserving unstable misalignments in ETGs. As these constitute a large fraction ($\approx0.76$) of the overall relaxation sample, this may help account for the morphological dichotomy in the instantaneous misalignment distributions (see Section~\ref{results: misalignment distributions}).

\subsubsection{Gas fraction dependence on relaxation timescales}\label{results: relaxations and gas fractions}

\begin{figure}
    \begin{center}
    \includegraphics[width=8.5cm]{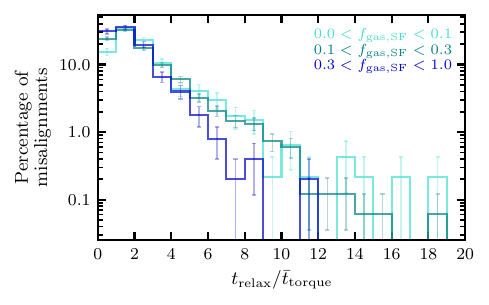}\\
    \caption{Normalised relaxation time distributions for relaxations with $t_{\mathrm{relax}}>0.2$~Gyr for a range of \gassf fractions. Percentages are given with respect to total relaxations within the given \gassf fraction sub-sample. Errors are given as Poisson uncertainties. We find a trend toward longer normalised relaxation times for smaller \gassf fractions with medians of $2.0t_{\rm{relax}}/\bar{t}_{\rm{torque}}$, $1.8t_{\rm{relax}}/\bar{t}_{\rm{torque}}$, and $1.4t_{\rm{relax}}/\bar{t}_{\rm{torque}}$ for increasing $f_{\rm{gas,SF}}$ sub-samples.}
    \label{plot: gassf ttorque}
    \end{center}
\end{figure}

In Figure~\ref{plot: gassf ttorque} we show the variation in normalised relaxation time for a range of \gassf fractions within $<r_{50}$, independent of morphology. Henceforth we exclude short-duration unstable misalignments close to the resolution limit (i.e.\,$t_{\mathrm{relax}}<0.2$~Gyr), due to the incompleteness and the larger uncertainties associated with relaxation timescales in this range. We retain a sample of 2614 relaxations, of which 466 lie within $0<f_{\rm{gas,SF}}<0.1$, 1646 within $0.1<f_{\rm{gas,SF}}<0.3$, and 502 within $0.3<f_{\rm{gas,SF}}<1$.

We see a trend toward longer-duration unstable misalignments for lower $f_{\rm{gas,SF}}$. This is seen clearly between the middle and highest $f_{\rm{gas,SF}}$ sub-samples (two-sample KS-test statistic $=0.13$ and p-value $=4.6\times10^{-5}$), and less strongly for the lowest and middle $f_{\rm{gas,SF}}$ sub-samples (two-sample KS-test statistic $=0.09$ and p-value $=5.5\times10^{-3}$). This is also reflected in the median normalised relaxation times, with medians of $2.0t_{\rm{relax}}/\bar{t}_{\rm{torque}}$, $1.8t_{\rm{relax}}/\bar{t}_{\rm{torque}}$, and $1.4t_{\rm{relax}}/\bar{t}_{\rm{torque}}$ for increasing $f_{\rm{gas,SF}}$ sub-samples. These results are unchanged when considering only \coco and \countercounter relaxations, and we conclude that bias from different relaxation path compositions is negligible. We expect these results to correlate with morphology because ETGs in \textsc{eagle} typically host lower star-forming gas fractions compared to LTGs \citep[e.g.][]{Casanueva2022}, echoing existing observational results \citep[see e.g.][]{Ruffa2019a, Ruffa2019b}.

We attribute the observed anti-correlation between $f_{\rm{gas,SF}}$ and $t_{\rm{relax}}/\bar{t}_{\rm{torque}}$ ratios to the lower gas disc angular momentum of gas-poor systems. These systems are affected more strongly by accretion onto the gas disc that may supply sufficient angular momentum through smooth accretion in order extend the lifetime of an unstable misalignment. If this smooth accretion is predominantly supplied by halo cooling (from a misaligned gaseous halo) we could expect longer normalised relaxation timescales in systems with 1) higher halo cooling (typically those with a higher halo mass) and 2) that have experienced more galaxy mergers and interactions (needed to misalign the halo). Both of these are preferentially found in ETGs. Indeed, galaxies in the range $t_{\rm{relax}}/\bar{t}_{\rm{torque}}\gtrsim6$ are predominantly gas-poor ETGs. 

It is worth noting that our strict criterion for relaxations creates a bias toward higher $f_{\rm{gas,SF}}$. This significantly reduces the number of gas-poor systems in our sample with $f_{\rm{gas,SF}}\lesssim0.05$, which may otherwise show even longer unstable misalignment durations. 

\subsubsection{Ongoing gas inflow rate dependence on relaxation timescales}\label{results: relaxations and inflow}

Ongoing accretion has been shown to extend the lifetime of an unstable misalignment beyond theoretical predictions \citep[e.g.][]{VoortDavis2015}. Here, we aim to verify whether this behaviour is also present in \textsc{eagle}.

We estimate the ongoing gas inflow rate by tracing gas particles entering a sphere with a radius of $2r_{50}$. This allows us to consider gas accretion onto gas discs that extend beyond our kinematic aperture ($r_{50}$). From visual inspection, we ignore the gas inflow over the first 0.3~Gyr post-formation, which sets a lower limit of $t_{\mathrm{relax}}>0.3$~Gyr. This allows us to largely ignore the initial gas accretion that may have formed the misalignment and focus on ongoing gas accretion. We time-average the gas inflow rate, $\dot{M}_{\mathrm{gas}}$, for the remainder of the relaxation. We limit our analysis to central galaxies because a significant proportion of unstable misalignments in satellite galaxies are likely affected by gas-stripping \citep[e.g. ram-pressure stripping;][]{Khim2021} rather than gas accretion. We use 1148 relaxations for which the central galaxy classification remains unchanged for the duration of $t_{\rm{relax}}$ and for which the ongoing gas inflow rates can be estimated. Of these, 286 have ongoing gas accretion rates of $\dot{M}_{\mathrm{gas}}<1$~M$_{\odot}$~yr$^{-1}$, 773 lie within $1<\dot{M}_{\mathrm{gas}}<10$~M$_{\odot}$~yr$^{-1}$, and 89 have higher ongoing accretion rates of $\dot{M}_{\mathrm{gas}}>10$~M$_{\odot}$~yr$^{-1}$.

\begin{figure}
    \begin{center}
    \includegraphics[width=8.5cm]{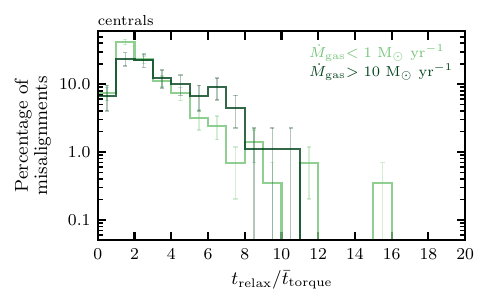}\\
    \caption{Normalised relaxation time distributions of central galaxies with gas inflow rates of $\dot{M}_{\mathrm{gas}}<1$~M$_{\odot}$~yr$^{-1}$ and $\dot{M}_{\mathrm{gas}}>10$~M$_{\odot}$~yr$^{-1}$, averaged 0.3~Gyr post-formation at $2r_{50}$. Percentages are given with respect to total relaxations within the inflow rate sub-sample. Errors are given as Poisson uncertainties. We find high gas inflow leads to longer normalised relaxation times than low inflow, with medians of $2.8t_{\mathrm{relax}}/\bar{t}_{\mathrm{torque}}$ and $2.0t_{\mathrm{relax}}/\bar{t}_{\mathrm{torque}}$, respectively.}
    \label{plot: inflow ttorque}
    \end{center}
\end{figure}

In Figure~\ref{plot: inflow ttorque} we show how the normalised relaxation time varies with the time-averaged ongoing gas inflow rate (where we have omitted the intermediate inflow range for clarity). Central galaxies with higher gas inflow rates show longer-lived unstable misalignments. We find median normalised relaxation times of $2.0t_{\mathrm{relax}}/\bar{t}_{\mathrm{torque}}$, $2.5t_{\mathrm{relax}}/\bar{t}_{\mathrm{torque}}$, and $2.8t_{\mathrm{relax}}/\bar{t}_{\mathrm{torque}}$ for $\dot{M}_{\mathrm{gas}}<1$~M$_{\odot}$~yr$^{-1}$ to $1<\dot{M}_{\mathrm{gas}}<10$~M$_{\odot}$~yr$^{-1}$ to $\dot{M}_{\mathrm{gas}}>10$~M$_{\odot}$~yr$^{-1}$, respectively. This effect is seen most clearly in the range $2\gtrsim t_{\mathrm{relax}}/\bar{t}_{\mathrm{torque}}\gtrsim8$. The difference between the lower and higher inflow galaxies result is significant (two-sample KS-test statistic $=0.16$ and p-value $=1.6\times10^{-4}$). Further splitting the sample among ETGs and LTGs reveals the significance of this result to be driven primarily by central ETGs rather than central LTGs (not shown). Averaging the gas inflow rate after a longer time since initial misalignment formation (e.g. $>0.5$~Gyr) strengthens the dichotomy between low- and high-inflow unstable misalignments.

These results suggest that galaxies experiencing higher ongoing gas accretion onto a misaligned gas disc can have their relaxation timescales enhanced. However, due to the resolution limitations of large-scale cosmological simulations, we caution against this result as a conclusive indicator of a relationship between enhanced relaxation times and ongoing inflow rates. 

\subsubsection{Central vs satellite galaxy dependence on relaxation timescales}\label{results: relaxations and occupation}

To investigate the effect of the environment on the relaxation distributions, we use 2470 relaxations for which the central or satellite galaxy classification remains unchanged for the duration of $t_{\rm{relax}}$. The application of this criterion returns 1699 centrals and 711 satellites.

We find a small trend toward longer normalised relaxation times in central galaxies compared to satellite galaxies with median relaxation times of $1.8t_{\rm{relax}}/\bar{t}_{\rm{torque}}$ and $1.6t_{\rm{relax}}/\bar{t}_{\rm{torque}}$, respectively (not shown).  Comparing the distributions of normalised relaxation times between central and satellite galaxies we find this difference is significant (two-sample KS-test statistic $=0.08$ and p-value $=2.0\times10^{-3}$). 

These results indicate that central galaxies appear to be more prone to additional physical processes extending the lifetimes of their misaligned gas compared to satellite galaxies. 

\subsubsection{Halo mass dependence on relaxation timescales}\label{results: relaxations and halo}

\begin{figure}
    \begin{center}
    \includegraphics[width=8.5cm]{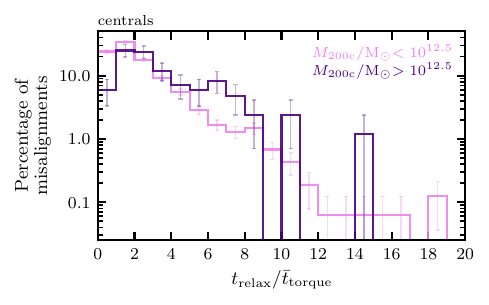}\\
    \caption{Normalised relaxation time distributions of central galaxies with different halo mass ranges for relaxations with $t_{\mathrm{relax}}>0.2$~Gyr. Percentages are given with respect to total relaxations within the halo mass sub-sample. Errors are given as Poisson uncertainties. We find a trend toward longer normalised relaxation times with larger halo masses, with medians of $1.7t_{\rm{relax}}/\bar{t}_{\rm{torque}}$ (low-mass) and $2.8t_{\rm{relax}}/\bar{t}_{\rm{torque}}$ (intermediate/high-mass). }
    \label{plot: halomass ttorque}
    \end{center}
\end{figure}

Finally, smooth accretion from misaligned halo cooling has been suggested as both a formation path of misalignments \citep[e.g.][]{Lagos2015}, and a means to extend the lifetimes of an existing unstable misalignment \citep[e.g.][]{Bryant2019}. Thus, we consider the effects of overall environment density (and by proxy, larger halo gas reservoirs) on the relaxation distribution by using the halo mass, $M_{\rm{200c}}$, as a proxy. These results are shown in Figure~\ref{plot: halomass ttorque}. 

We limit ourselves to central galaxies for this analysis as gas cooling from a hot halo will primarily accrete onto the central galaxy \citep[e.g.][]{Fabian84, Negri2014, Gaspari17}.  We consider two mass ranges broadly corresponding to low-mass halos ($M_{\rm{200c}}<10^{12.5}$ M$_{\odot}$) and intermediate to high-mass halos ($M_{\rm{200c}}>10^{12.5}$ M$_{\odot}$). This provides a sample of 1699 relaxations with sub-sample sizes of 1615 (low-mass) and 84 (intermediate/high-mass). Because we find only 2 instances of central cluster galaxies (assumed to be $M_{\rm{200c}}>10^{14}$ M$_{\odot}$) among our sample, we only distinguish between low- and intermediate-mass environments. 

We find a strong and significant difference between the two halo masses considered (two-sample KS-test statistic $=0.29$ and p-value $=1.5\times10^{-6}$) showing a positive correlation between halo mass and the normalised relaxation time (in terms of $\bar{t}_{\rm{torque}}$). This is seen in both the overall shape of the distributions in Figure~\ref{plot: halomass ttorque} and in the median relaxation times which correspond to $1.7t_{\rm{relax}}/\bar{t}_{\rm{torque}}$ (low-mass) and $2.8t_{\rm{relax}}/\bar{t}_{\rm{torque}}$ (intermediate/high-mass). Low-mass halos are dominated by short-duration normalised relaxation times, suggesting processes that preserve unstable misalignments (e.g. smooth accretion) are rare in this mass regime. On the other hand, intermediate/high-mass halos have a considerably flatter distribution and dominate the long-duration normalised relaxation timescales beyond $\gtrsim4t_{\rm{relax}}/\bar{t}_{\rm{torque}}$. This behaviour becomes more pronounced when considering central galaxies with $M_{\rm{200c}}>10^{13.5}$ M$_{\odot}$ (not shown). 

Although not shown, we also find a strong positive relationship between normalised relaxation times (in terms of $\bar{t}_{\rm{torque}}$) and stellar mass within $2r_{50}$. $t_{\rm{relax}}/\bar{t}_{\rm{torque}}$ increases from 1.6 to 2.5 to 3.7 for mass ranges of $M_{\rm{*}}<10^{10}$~M$_{\odot}$, $10^{10}$~M$_{\odot}$ $<M_{\rm{*}}<10^{11}$~M$_{\odot}$, and $M_{\rm{*}}>10^{11}$~M$_{\odot}$, respectively. Again, these results are largely unchanged when considering only \coco and \countercounter relaxation paths. 

Together, these results point toward halo cooling \citep[from misaligned halos; see e.g.][]{Hill2021} being an important driver of long-duration misalignments in higher-mass systems in \textsc{eagle}. However, we stress that a more thorough analysis needs to be done to confirm the significance of accretion from misaligned halo cooling as a driver of misalignments within \textsc{eagle} and similar cosmological simulations.

\subsection{Incidence of mergers with misalignment formation}\label{results: misalignment origins}

In this section, we briefly explore the incidence of mergers with the formation of unstable misalignments. For this analysis, we consider galaxies with $M_{*}\gtrsim10^{10}$~M$_{\odot}$ in order to more accurately resolve minor mergers in the merger tree. This gives us a sample of 718 relaxations, with statistics given in Table~\ref{tab: median relaxation times 1010}. The behaviour of median \trelax values for ETGs and LTGs are comparable to those of our original sample, with a trend toward a larger and flatter distribution of $t_{\rm{relax}}/\bar{t}_{\rm{dyn}}$ and $t_{\rm{relax}}/\bar{t}_{\rm{torque}}$ as expected from our results in Section~\ref{results: relaxations and halo}. Alongside higher gas fractions and SFRs, galaxies at $z\approx1$ experience more frequent galaxy interactions and mergers compared to $z\approx0$ counterparts \citep[][]{Fensch17, Pfeffer2023}. As such, we further sub-divide our sample between high-$z$ ($0.35<z<1.00$) and low-$z$ ($z<0.35$) with sub-samples of 449 and 269 relaxations, respectively. These redshift ranges correspond to approximately equal temporal ranges of $\sim4$~Gyr.

The incidence of mergers and the formation of unstable misalignments will be sensitive to the time window over which we associate these two events. We search for mergers with stellar mass ratios of $\mu_{\rm{*}}>0.1$ within windows of 0.4~Gyr, 0.6~Gyr, and 1.0~Gyr centred on the first snipshot to enter the unstable [$20^{\circ}-160^{\circ}$] regime. As galaxies tend to become unstably misaligned quickly, this snipshot overwhelmingly coincides with the first unstable misalignment [$30^{\circ}-150^{\circ}$]. Focusing primarily on the 0.6~Gyr window size, we aim to capture unstable misalignment formations that are explicitly associated with a direct merger origin (such as the first unstable misalignment in Galaxy B in Figure~\ref{plot: evolutions}) and exclude unstable misalignment formations associated with an accretion event from an initial fly-by (such as Galaxy A in Figure~\ref{plot: evolutions}). By setting this window to 0.6~Gyr we search for mergers within $\approx2$ snipshots on either side of the kinematic instability. This gives us a small degree of flexibility with regard to the exact timing of the merger according to the existing merger trees, and gives the gas a short window within which to settle and form an unstable misaligned disc. Because minor merger signatures are typically visible for $\sim0.5$~Gyr in observational studies \citep[e.g.][]{Lotz10}, we reason that any unstable misalignments associated with mergers in our sample would likely be classified similarly in optical counterparts. 

\begin{table}
 \caption{Statistics on the median and standard deviation (SD) of relaxation timescales for a sub-sample of $M_{*}>10^{10}$ M$_{\odot}$ galaxies, expressed for the total sample and by morphology (e.g. ETG $\rightarrow$ ETG). Relaxation timescales are given in terms of~Gyr, and as normalised relaxation times in terms of $\bar{t}_{\rm{dyn}}$, and $\bar{t}_{\rm{torque}}$.}
     \label{tab: median relaxation times 1010}
 \begin{tabular}{p{2.1cm}>{\centering}p{0.65cm}>{\centering}p{0.5cm}>{\centering}p{0.65cm}>{\centering}p{0.5cm}>{\centering}p{0.65cm}>{\centering\arraybackslash}p{0.5cm}}
 \toprule 
    \multirow{2}{*}{\centering Relaxation timescales} &
    \multicolumn{2}{c}{All} &
    \multicolumn{2}{c}{LTG $\rightarrow$ LTG} &
    \multicolumn{2}{c}{ETG $\rightarrow$ ETG} \\
 \cmidrule(lr){2-3} \cmidrule(lr){4-5} \cmidrule(lr){6-7}
    &
    Median & SD &
    Median & SD &
    Median & SD \\
 \midrule
    $t_{\rm{relax}}$ (Gyr)  & 0.38 & 0.62 &   0.26 & 0.34 &    0.48 & 0.72 \\
    $t_{\rm{relax}}/\bar{t}_{\rm{dyn}}$   & 3.47 & 6.03 &   2.71 & 4.57 &    3.95 & 6.75 \\
    $t_{\rm{relax}}/\bar{t}_{\rm{torque}}$  & 1.87 & 2.86 &   1.59 & 2.54 &    1.85 & 3.04 \\
 \bottomrule
 \end{tabular}
\end{table}

\begin{figure*}
    \includegraphics[width=17.5cm]{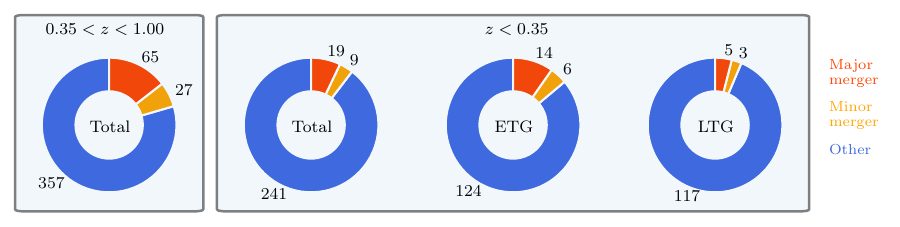}\\
    \caption{Pie charts visualizing the number of unstable misalignments associated with major and minor mergers for $0.35<z<1.00$ and $z<0.35$, with the latter further split between ETGs and LTGs, all with $M_{*}>10^{10}$ M$_{\odot}$. Overall, we find $\approx10.4$~per~cent ($\approx20.5$~per~cent) of our relaxation sample associated with mergers within redshift ranges $z<0.35$ ($0.35<z<1.00$).}
    \label{plot: misalignment origins}
\end{figure*}

In Figure~\ref{plot: misalignment origins} we visualise the fraction of unstable misalignments associated with major and minor mergers within a 0.6~Gyr window for sub-samples of high-$z$ and low-$z$ relaxing misalignments. For our low-$z$ sub-sample, we additionally show the merger dependence on the morphology \textit{before} any misalignment (e.g. ETG $\rightarrow$ LTG relaxations are classified under `ETG'). 

At low-$z$, $\approx10.4$~per~cent of our sample is associated with a merger origin, with major mergers ($\approx7.1$~per~cent) dominating this fraction over minor mergers ($\approx3.3$~per~cent) in both morphologies. Of these, unstable misalignments in ETGs show higher merger-origin fractions ($\approx13.9$~per~cent) over LTGs ($\approx6.4$~per~cent) which may be due to ETGs being more susceptible to misalignments from stochastic external accretion owing to lower \gassf fractions. Changing the window size does not change the relative proportions of major and minor mergers nor the higher incidence in ETGs, but does affect the incidence of mergers with $\approx4.4$~per~cent and $\approx14.5$~per~cent (both within $z<0.35$) for windows of size 0.4~Gyr and 1.0~Gyr, respectively. Among our high-$z$ sub-sample, we find a higher incidence of mergers $\approx20.5$~per~cent) though they remain sub-dominant in misalignment formation. Together with our low-$z$ sub-sample, we find an incidence of $\approx16.7$ for relaxations in $0<z<1$. These results imply that major mergers are more effective at driving unstable misalignments than minor mergers, as the latter may lack sufficient gas to meaningfully change the angular momentum of the in-situ gas. 

We conclude that galaxy mergers are not the primary driver behind stellar-gas misalignments in \textsc{eagle} galaxies, but they may have been more important at higher redshifts. This once again points toward significant contributions of other processes that form unstable misalignments, which are in agreement with results from \citet{Casanueva2022}. However, given that misaligned halos are often associated with galaxy encounters and mergers \citep[e.g.][]{BettFrenk12}, external processes may remain dominant in misalignment formation but these may only be weakly connected temporally.


\subsection{Discussion and comparison to existing results}\label{discussion:} 

One of the key issues in this field in recent years has been the inability of current models to accurately reproduce the observed misalignment distribution at $z\sim0.1$ \citep[e.g.][]{Lagos2015, DavisBureau2016}. In this section, we put our results in the context of the proposed solutions to this discrepancy.

As explored in detail in \citet{DavisBureau2016}, assuming misalignments to be formed primarily through mergers alongside theoretical relaxation timescales (in the absence of smooth gas accretion), analytical models significantly overproduce the fraction of counter-rotating (relaxed) ETGs and under-produce the fraction of unstable misaligned (un-relaxed) ETGs \citep[][]{DavisBureau2016}. While shorter gas depletion timescales can be invoked to reproduce observed distributions, such that a misaligned disc is depleted before it can relax, this solution is not favoured. This is because it requires both a significantly higher merger rate ($\sim5-10$ times larger than predicted from $\Lambda$CDM) to supply frequent misaligned gas alongside a higher efficiency of processes to deplete it (e.g. increased AGN activity or star-formation efficiency) that far exceeds values measured for typical ETGs.

One solution presented by \citet{DavisBureau2016} suggests that the relaxation timescales of many ETGs may be significantly underestimated due to significant smooth accretion onto unstable misaligned gas discs. Assuming mergers to be the dominant driver of misalignments, relaxation timescales $\sim10$ times higher ($\sim20-80 t_{\mathrm{dyn}}$) could broadly reproduce the observed ETG misalignment angle distributions. This behaviour was studied by \citet{VoortDavis2015} using the zoom-in \textsc{fire} simulations. Following a gas-rich merger, an unstable misaligned gas disc persisted for $\approx2.1$~Gyr attributed to low levels of smooth accretion onto the misaligned disc from tidal tails. This significantly extended the lifetime of the misalignment for the first $\approx1.8$~Gyr before a decrease in accretion rate led to an eventual decoupling between the misaligned disc and accretion flow, followed by a relaxation. 

An alternative solution suggests that there may be additional processes forming misalignments besides mergers. For instance, galaxies that are already gas-poor \citep[such as red-sequence ETGs; ][]{Davis2022} have a lower angular momentum threshold needed to misalign the gas disc. This may make these systems more susceptible to becoming misaligned from other sources of misaligned accretion \citep[e.g.][]{Lagos2018a, Khim2021, Casanueva2022}. Indeed, recent results from semi-analytical models and simulations have shown the formation pathways for misalignments to be significantly more diverse than previously assumed in earlier observational studies. Using the \textsc{galform} semi-analytical model \citep{Cole00} and assuming mergers to be the only source of misalignments, \citet{Lagos2015} found misalignments in only $\sim2-5$~per~cent of ETGs at $z=0$. However, when additionally considering isotropic smooth accretion from halo cooling, the fraction of misaligned ETGs increased to $\sim46$~per~cent of ETGs bringing this value significantly closer to observational results \citep[e.g.][]{Davis2011, Bryant2019, Ristea2022}. More recently, \citet{Khim2021} use the \textsc{horizon-agn} simulation to investigate the origins and timescales of misalignments for a sample of $\approx27,900$ galaxies. They report four main formation paths for misalignments: mergers ($\approx35$~per~cent), gas stripping and interactions with the environment ($\approx23$~per~cent), interactions with other galaxies ($\approx21$~per~cent), and secular evolution such as filament accretion or misaligned halo cooling ($\approx21$~per~cent). 

Overall, our results tend to be in agreement with the presence of additional drivers of misalignments, rather than a dominant population of significantly longer-lived unstable misalignments in ETGs. From our results in Section~\ref{results: relaxation timescales with morphology}, we find that the vast majority of unstable misalignments in ETGs relax within $\sim6-8\bar{t}_{\mathrm{dyn}}$. These relaxations do not show substantially enhanced relaxation times compared to LTGs once timescales are normalised with respect to $\bar{t}_{\mathrm{torque}}$. While we do see a population of fairly long-duration unstable misalignments with $\sim1-2$~Gyr \citep[of comparable duration to][]{VoortDavis2015}, and a population with enhanced relaxation timescales of $>3\bar{t}_{\mathrm{torque}}$ beyond theoretical timescales \citep[][]{Tohline1982, LakeNorman1983}, these are not dominant among our sample. Furthermore, the relatively low incidence of mergers ($\approx10-21$~per~cent for low-$z$ and high-$z$, respectively) with the formation of unstable misalignments in ETGs suggests misalignments form through diverse formation pathways and may occur more frequently than previously assumed. Thus, at least in \textsc{eagle}, we conclude that a population of long-duration unstable misalignments formed through mergers in ETGs, as suggested by \citet{DavisBureau2016}, is not the main driver behind the misalignment distributions at $z\sim0.1$. 

In qualitative terms, our relaxation timescales agree with those investigated in \textsc{horizon-agn} by \citet{Khim2021}. Using a sample of misaligned ($>30^{\circ}$) galaxies and characterising relaxation timescales as a population decay, they find longer characteristic decay timescales in ETGs ($\approx2.34$~Gyr) compared to LTGs ($\approx0.75$~Gyr). This gives ETGs a $\approx3.4$ times longer misalignment duration. By tracing the longevity of individual unstable misalignments in \textsc{eagle}, we tend to find on average shorter relaxation timescales overall. However, we also find longer relaxation times in ETGs ($\approx0.51$~Gyr) compared to LTGs ($\approx0.38$~Gyr). Besides underlying differences in approach (e.g. use of a decay timescale) and the different mass ranges considered, the longer relaxation times found in \textsc{horizon-agn} may be due to numerous reasons. Firstly, \textsc{horizon-agn} contains a larger number of $M_*>10^{10}$ M$_{\odot}$ galaxies ($\approx28,000$) compared to \textsc{eagle} ($\approx3600$) at $z\sim0$. These galaxies typically reside in denser environments in \textsc{horizon-agn} compared to \textsc{eagle} \citep[e.g.][]{Khim2020}. This, in turn, may increase relaxation times due to increased galaxy encounters and smooth accretion events \citep{Khim2021}. Secondly, as the time intervals between successive simulation outputs is about twice as long in \textsc{horizon-agn} compared to \textsc{eagle}, galaxies in our sample are required to meet our sample criteria over a greater number of consecutive snapshots. The lower time resolution in \textsc{horizon-agn} may also obfuscate many shorter-duration misalignments that we are able to resolve. Thirdly, \citet{Khim2021} make use of projected 2D misalignment angles. While we found no substantial changes in the shape of the misalignment distribution at $z=0.1$ between 2D and 3D misalignment angles (see Figure~\ref{plot: misalignment distributions}), it is clear that the use of 3D misalignment angles is preferred when characterising relaxation timescales with respect to theoretical predictions (see Section~\ref{theory:}). Finally and most importantly, while we aim to measure the relaxation timescales of \textit{unstable} misalignments ($30^{\circ}-150^{\circ}$), \citet{Khim2021} measure the longevity of misaligned galaxies ($>30^{\circ}$). As a consequence, galaxies that spend considerable time in the stable counter-rotating regime (e.g. Galaxy B in Figure~\ref{plot: evolutions}) would have characteristic misalignment timescales ($\approx4.5$~Gyr) far exceeding the relaxation timescales that we measure ($\approx0.35$~Gyr and $\approx2.55$~Gyr). 

\citet{Bryant2019} highlight that it is difficult to achieve the observed dichotomy between ETG and LTG misalignment distributions at $z\approx0.1$ through differences in theoretical relaxation timescales. Using a slightly modified expression for \ttorque and assuming a standard ellipticity of 0.2 and 0.8 for ETGs and LTGs respectively, they find a theoretically maximum $2.7\times$ longer relaxation time in ETGs. Thus, if morphology is the primary driver of the observed dichotomy and assuming unstable misalignments are formed at the same rate in ETGs and LTGs, we should expect $<2.7\times$ more misaligned ETGs than LTGs. Further considering variations in the typical size and radial velocity of gas discs and taking ellipticity values from a full realistic population of galaxies, this ratio drops to $\lesssim1.4$. In \textsc{eagle}, we find a similarly low ratio of $\approx1.3\times$ ($\approx1.6\times$ for \coco relaxation paths) between median relaxation times of ETGs and LTGs. Crucially, these ratios are significantly lower than the observed ratio of misaligned gas between ETGs and LTGs ($15\pm7$) in SAMI and similar studies. Likewise, we find a misaligned \gassf ratio of $6.2\pm1.2$ between ETGs and LTGs (see Figure~\ref{plot: misalignment distributions}). Thus, we concur with \citet{Bryant2019} that differences in theoretical relaxation timescales between ETGs and LTGs are not the dominant driver behind the observed misalignment distributions. Instead, we suggest that a higher frequency of misalignment events in ETGs is likely the dominant driver behind the observed misalignment distributions.

We find a relatively low incidence of mergers coinciding with the formation of unstable misalignments ($\approx11-22$~per~cent) over $0<z<1$, highlighting the presence of additional processes forming unstable misalignments, especially within $z<0.35$. These results are also in qualitative agreement with recent work in cosmological simulations \citep[e.g.][]{Starkenburg2019, Khim2021, Casanueva2022}, as well as semi-analytical models \citep[e.g.][]{Lagos2015}. For instance, in earlier work using \textsc{eagle}, \citet{Casanueva2022} report only $\approx4$~per~cent of misalignments ($>30^{\circ}$) coinciding with mergers and highlight the importance of gas depletion (possibly through outflows or stripping) to form misalignments through galaxy interactions and encounters. Quantitative differences are likely caused by our different methods (as we use \textsc{eagle} snipshots and utilise a time window within which to identify a merger-driven origin). The mass range of galaxies considered is also different, with \citet{Casanueva2022} including many dwarf galaxies ($M_{*}>10^{9}$~M$_{\odot}$) that may be located in isolated environments with few merger opportunities. 

Our results are also in relatively good qualitative agreement with \citet{Khim2021} who found a major (minor) merger-incidence of $\sim16$~per~cent ($\sim15$~per~cent) at $z\approx0$ using a modified major and minor merger ratio definition of $\mu_{*}>0.25$ and $0.25>\mu_{*}>0.02$, respectively. For comparison, we find a major (minor) merger-incidence of $\approx7.1$~per~cent ($\approx3.3$~per~cent) for galaxies within $z<0.35$. Quantitative differences likely arise from the inclusion of smaller merger ratios in their method alongside the aforementioned higher environment density of galaxies in \textsc{horizon-agn}, prompting more frequent mergers and galaxy encounters. Our results also show some agreement with recent observational results in SAMI within $z<0.1$, with a likely merger-origin associated with $\approx14$~per~cent of misalignments in galaxies with a mass range of $10^{9}\lesssim M_{*}\lesssim10^{11.5}$ M$_{\odot}$ \citep[][]{Ristea2022}.

Finally, we find a qualitative agreement with \citet{Khim2021} on the relation between long-lived misalignments and halo mass as a proxy for environment density. While we find no significant difference between the \trelax of low and medium/high-mass halos, we do find a strong positive correlation between $t_{\mathrm{relax}}/\bar{t}_{\mathrm{torque}}$ and halo mass. We also find a positive correlation between $t_{\mathrm{relax}}/\bar{t}_{\mathrm{torque}}$ and higher ongoing gas accretion rates among central galaxies. Alongside \citet{Khim2021}, we attribute the positive correlation between halo mass and longer-lived unstable misalignments to both a higher incidence of interactions and gas stripping events in group environments, as well as higher quantities of halo cooling (and resulting gas inflow) with increasing halo mass \citep[e.g.][]{Correa2018haloacc, MitchellSchaye2022halo}. For the latter, an unstable misalignment in a more massive halo may be preserved through significant misaligned smooth gas accretion, provided the gaseous halo is misaligned. For instance, stellar-gas misalignments were found to be more common in galaxies with misaligned halos in the IllustrisTNG simulation \citep{Duckworth2020a}. The combination of lower \gassf masses and a higher frequency of halo angular momentum decoupling likely makes ETGs more prone to unstable misalignments from processes besides mergers. These conclusions are also in agreement with those by \citet{Casanueva2022}.

We note that a possible limiting factor in our analysis is due to our strict relaxation sample criteria. As a natural consequence of requiring consecutive criteria-meeting snipshots for the duration of the unstable misalignment, the chances of us excluding a relaxation increase significantly for longer-duration relaxations. Long-duration unstable misalignments in ETGs are especially vulnerable to this, given that typical ETGs in \textsc{eagle} have both lower \gassf particle counts and host longer-duration unstable misalignments. Similarly, there is likely an additional population of unstable misalignments in galaxies with ongoing mergers. For example, \citet{Khim2021} find misalignments driven by group interactions to have the longest duration. However, as these relaxations contain snipshots that do not meet our sample criteria, or fail to relax before merging, we reason that any estimate of the relaxation time in these galaxies is unreliable. 

As highlighted in Appendix~\ref{Appendix: latency time}, we find no substantial difference in the number of long-lived ($t_{\mathrm{relax}}>0.1$~Gyr) unstable misalignments identified through use of a 0, 0.1 or 0.2 Gyr latency time. Thus, we rule out the possibility of missing a significant number of long-lived unstable misalignments due to our choice of $t_{\mathrm{lat}}=0.1$~Gyr. Finally, our sample redshift window of $0<z<1$ ($\approx8$~Gyr) naturally limits the number of complete $t_{\mathrm{relax}}\gtrsim20t_{\mathrm{dyn}}$ misalignments we can measure from formation to relaxation, with $t_{\mathrm{relax}}\approx40t_{\mathrm{dyn}}$ being an approximate upper limit. However, considering the limitations imposed by our relaxation sample criteria, the fact we still find a moderate population of long-lived misalignments suggests that misalignment-preserving processes are nevertheless present and that these may be more widespread than previously assumed.


\section{Conclusions}\label{conclusions:} 

In this work, we have used the \textsc{eagle} cosmological simulations to investigate the longevity of stellar-gas kinematic misalignments
in galaxies with a stellar mass $>10^{9.5}$ M$_{\odot}$. We use samples of $\approx5400$ galaxies for each snapshot to investigate the distribution and variation in the fraction of misaligned systems ($\psi_{\mathrm{3D}}>30^{\circ}$) between $0<z<1$. We then construct a sample of 3154 misalignment relaxations for which we can reliably estimate the time taken for an unstable misalignment ($30^{\circ}<\psi_{\mathrm{3D}}<150^{\circ}$) to settle into the kinematically stable regime ($\psi_{\mathrm{3D}}<20^{\circ}$ or $\psi_{\mathrm{3D}}>160^{\circ}$). For each relaxation, we compare the relaxation time ($t_{\mathrm{relax}}$) to the dynamical timescale of the disc ($\bar{t}_{\mathrm{dyn}}$) and to the expected relaxation timescale from theoretical predictions (i.e. the torquing timescale, $\bar{t}_{\mathrm{torque}}$). Using this sample, we established the distribution of relaxation timescales for a variety of galaxy properties including morphology, star-forming gas fraction, and halo mass. Next, we briefly investigated the correlation between the incidence of mergers and the formation of unstable stellar-gas kinematic misalignments. Finally, we compare our results to theoretical predictions and interpret these in the context of driving the observed misalignment distribution at $z\approx0.1$. 

Our results can be summarised as follows:

\begin{enumerate}
\item For both ETGs and LTGs, the fraction of 3D misalignments ($>30^{\circ}$) increases slightly toward lower redshifts, driven by the build-up of long-lasting counter-rotating ($>150^{\circ}$) systems. By $z=0.1$, we find counter-rotating gas discs present in $\approx6$~per~cent of our sample. 

\item The majority of unstable misalignments relax on short timescales, with median relaxation times of $t_{\rm{relax}}\approx0.5$~Gyr $\approx3.2\bar{t}_{\rm{dyn}}\approx1.4\bar{t}_{\rm{torque}}$. These indicate the majority of unstable misalignments do not experience significantly enhanced relaxation timescales of $\sim20\bar{t}_{\rm{dyn}}$ due to continued smooth accretion, as suggested by \citet{DavisBureau2016} to explain the shape of the ETG misalignment distribution at $z\sim0.1$. 

\item ETGs take $\approx1.3$ longer to relax compared to LTGs. However, accounting for differences in dynamical timescales and mass distributions between morphologies, ETGs show no significant preference over LTGs toward longer-lived unstable misalignments.

\item We consider long-duration relaxations as those with $t_{\rm{relax}}>3\bar{t}_{\rm{torque}}$ and find that these constitute $\approx20$~per~cent of our sample. Galaxies with these long-duration unstable misalignments tend to have lower star-forming gas fractions, above-average stellar mass, and  higher ongoing gas inflow rates. 

\item We find a positive correlation between halo mass and normalised relaxation timescales, with central galaxies residing in intermediate/high-mass environments ($M_{\rm{200c}}>10^{12.5}$ M$_{\odot}$) showing clear signs of processes preserving unstable misalignments beyond theoretical predictions. We speculate some of these extended lifetimes may be driven by the cooling of hot gas from a misaligned halo. 

\item Among galaxies with $M_{*}>10^{10}$ M$_{\odot}$, mergers coincide with the formation of an unstable misalignment in $\approx10.4$~per~cent of cases at $z<0.35$. Major mergers are more effective at forming unstable misalignments and dominate the formation of unstable misalignments ($\approx7.1$~per~cent) compared to minor mergers ($\approx3.3$~per~cent). Mergers contribute toward the formation of unstable misalignments more in ETGs ($\approx13.9$~per~cent) than in LTGs ($\approx6.4$~per~cent). At $0.35<z<1.00$, the incidence of mergers that coincide with misalignment formation increases to $\approx20.5$~per~cent.
\end{enumerate}

Taken as a whole, these results point toward a more complex picture regarding the origins and drivers of kinematic misalignments in galaxies than previously considered. Given the relatively low incidence of mergers with misalignment formation, alongside only moderately extended misalignment lifetimes, misalignments are likely occurring more frequently in ETGs given the smaller gas discs in these systems are more susceptible to perturbations. However, with our results pointing toward significant contributions of smooth accretion from the cooling of a hot misaligned halo, we likely still require galaxy interactions to misalign the halo in the first place. In this sense, misalignments are still formed primarily through external processes, but they are not necessarily temporally connected. Thus, the star-forming gas replenishment mechanisms in galaxies likely occur through a diverse source rather than a prevalence toward gas-rich minor mergers (as suggested for red-sequence ETGs).

More work needs to be done in this field to fully understand the drivers of these misalignments to discern the main ISM replenishment mechanisms of ETGs. Future observational surveys with e.g. ALMA and JWST alongside state-of-the-art cosmological hydrodynamical simulations with a multiphase ISM, such as FIREbox \citep[][]{Feldmann2023} and the upcoming \textsc{colibre} (Schaye et al., in prep; Chaikin et al. in prep), offer an opportunity to explore these unique systems in more detail.


\section*{Acknowledgements}

We thank the anonymous referee for constructive comments and recommendations. We acknowledge support from the UK Science Technologies and Facilities Council (STFC) for PhD studentship funding and support for TAD through grant ST/W000830/1. We acknowledge the Virgo Consortium for making their simulation data available and thank Robert Crain and Liverpool John Moores University for providing the \textsc{eagle} snipshots. The \textsc{eagle} simulations were performed using the DiRAC-2 facility at Durham, managed by the ICC, and the PRACE facility Curie based in France at TGCC, CEA, Bruy\`eres-le-Ch\^atel. FvdV is supported by a Royal Society University Research Fellowship (URF\textbackslash R1\textbackslash 191703 and URF\textbackslash R\textbackslash 241005).

\section*{Data Availability}

The \textsc{eagle} simulations and database are publicly available at: http://icc.dur.ac.uk/Eagle/database.php. Detailed guides to simulation parameters, models, and how to access and query data can be found in \citet{EAGLEparticleRef} and \citet{Mcalpine2016}. \textsc{eagle} "snipshots" are available on request from members of the \textsc{eagle} team.


\bibliographystyle{mnras}
\bibliography{library_paper1} 



\appendix

\section{Variation of latency time on our unstable misalignment relaxation sample}\label{Appendix: latency time}

The requirement of a latency time in our relaxation time method is a compromise between extracting many short relaxations (that may belong to the same overall relaxation) or extracting fewer but longer-duration relaxations (in which we merge many of these small-duration unstable misalignments). Our latency time ($t_{\rm{lat}}=0.1$ Gyr) was chosen by manual inspection of unstable misalignment evolution to reflect this compromise. 

Varying this latency time has several effects. Firstly, by increasing $t_{\rm{lat}}$, we remove an increasing fraction of short-duration relaxations as successive relaxations become merged into longer relaxations. Secondly, this also significantly reduces our sample size by requiring a longer time gap between the next unstable misalignment, as well as requiring a larger window of snipshots over which to meet our sample criteria. The interplay between these selection effects sets the overall relaxation time distribution.

We explore the selection effects of varying the latency time between $0<t_{\rm{lat}}<0.5$~Gyr, with statistics shown in Table~\ref{tab: relaxation sample selection effects}. While increasing $t_{\rm{lat}}$ comes at a significant cost to our overall sample size, we find no significant change in the overall shape of our relaxation timescale distribution besides an increase in the relative fraction of $t_{\rm{relax}}<0.3$~Gyr relaxations for $t_{\rm{lat}}=0$~Gyr and subsequently decreased median relaxation time. As expected, this is driven primarily by substantial increase in the relative number of short-duration unstable misalignments rather than a significant decrease in the number of long-duration unstable misalignments. We therefore rule out the possibility of missing any substantial population of long-duration unstable misalignments due to variations in $t_{\rm{lat}}$.

\begin{table}
 \caption{Variation of the number of relaxations in our sample (N), median relaxation time ($t_{\rm{relax}}$), and fraction of unstable misalignments with long relaxation times ($t_{\rm{relax}}>1$~Gyr) in our relaxation sample from $0<z<1$ with respect to varying latency time ($t_{\rm{lat}}$) between 0~Gyr and 0.5~Gyr.}
     \label{tab: relaxation sample selection effects}
 \begin{tabular}{p{0.9cm}>{\centering}p{0.6cm}>{\centering}p{1.1cm}>{\centering}p{1.9cm}>{\centering\arraybackslash}p{1.9cm}}
 \toprule 
    \multirow{2}{*}{$t_{\rm{lat}}$ (Gyr)} &
    \multirow{2}{*}{N} &
    \multirow{2}{*}{$t_{\rm{relax}}$ (Gyr)} &
    \multicolumn{2}{c}{$t_{\rm{relax}}>1$ Gyr relaxations} \\
 \cmidrule(lr){4-5}
    & & &
    count & fraction \\
 \midrule
    0   &   6343    &   0.36    &   681   &   0.104   \\
    0.1 &   3154    &   0.50    &   635   &   0.200   \\
    0.2 &   1983    &   0.55    &   535   &   0.270   \\
    0.5 &    865    &   0.55    &   267   &   0.309   \\
 \bottomrule
 \end{tabular}
\end{table}


\bsp	
\label{lastpage}
\end{document}